\documentclass[twocolumn]{aastex62}

\hypersetup{linkcolor=red,citecolor=blue,filecolor=cyan,urlcolor=magenta}
\usepackage{natbib}
\usepackage{amsthm}
\usepackage{amssymb}
\usepackage{amsmath}
\usepackage{bm}

\received{\today}
\revised{, 2020}
\accepted{, 2020}
\submitjournal{ApJ}

\shorttitle{Ion-Ion instabilities: Effects of Suprathermal Populations}
\shortauthors{Shaaban et al.}

\begin{document}

\title{Electromagnetic Ion-Ion Instabilities in Space Plasmas: Effects of Suprathermal Populations}

\correspondingauthor{S.M.Shaaban}
\email{s.m.shaaban88@gmail.com}

\author[0000-0003-0465-598X]{S.M.Shaaban}
\affil{Centre for Mathematical Plasma Astrophysics, KU Leuven, Celestijnenlaan 200B, B-3001 Leuven, Belgium}
\affiliation{Theoretical Physics Research Group, Physics Dept., Faculty of Science, Mansoura University, 35516, Mansoura, Egypt}

\author[0000-0002-8508-5466]{M. Lazar}
\affiliation{Centre for Mathematical Plasma Astrophysics, KU Leuven, Celestijnenlaan 200B, B-3001 Leuven, Belgium}
\affiliation{Institut f\"ur Theoretische Physik, Lehrstuhl IV: Weltraum- und Astrophysik, Ruhr-Universit\"at Bochum, D-44780 Bochum, Germany}

\author[0000-0001-8134-3790]{R.A.L{\'o}pez}
\affiliation{Departamento de Física, Universidad de Santiago de Chile, Casilla 307, Santiago, Chile}

\author[0000-0002-1743-0651]{S. Poedts}
\affiliation{Centre for Mathematical Plasma Astrophysics, KU Leuven, Celestijnenlaan 200B, B-3001 Leuven, Belgium}

\begin{abstract}

In collision-poor plasmas from space, three distinct ion-ion instabilities can be driven by the proton beams streaming along the background magnetic field: left-hand resonant,
non-resonant, and right-hand resonant instabilities. These instabilities are in general investigated considering only idealized proton beams with Maxwellian velocity distributions, and ignoring the implications of suprathermal populations, usually reproduced by the Kappa power-laws. Moreover, the existing theories minimize the kinetic effects of electrons, assuming them isotropic and Maxwellian distributed. In an attempt to overcome these limitations, in the present paper we present the results of an extended investigation of ion-ion instabilities, which show that their dispersion and stability properties (e.g. growth rates, wave frequencies, and the unstable wave numbers) are highly sensitive to the influence of suprathermal populations and anisotropic electrons. These results offer valuable explanations for the origin of the enhanced low-frequency  fluctuations, frequently observed in space plasmas and associated with proton beams. 

\end{abstract}


\section{Introduction} \label{sec:1}
Proton beams are ubiquitous in space plasmas, and have been reported by interplanetary missions for different, but more or less specific conditions in the solar wind \citep{Marsch1982, Leubner2004a}, bow shocks \citep{Paschmann1981, Eastwood2005}, interplanetary shocks \citep{Vinas1984, Lario2019}, coronal mass ejections (CMEs) \citep{Marsch2009}, cometary environments \citep{Neugebauer1989, Behar2018}, and near-Earth portion of the plasma sheet boundary layer \citep{Takahashi1988, Birn2020}.  Guided by the magnetic field lines these beams may provide information on the young solar wind acceleration and heating close to the Sun, and about the interplanetary magnetic field topology \citep{Gosling2005, Alterman2018}. Measured in-situ, proton velocity distribution functions often can be represented in terms of two counter-moving populations, namely, a dense thermal core and tenuous suprathermal beam \citep{Marsch1982, Goldstein2000, Marsch2009}. Suprathermal beams enhance the high-energy tails of the observed velocity distributions \citep{Formisano1973, Feldman1973, Goodrich1976, Marsch1982, Leubner2004a, Leubner2004b, Lario2019}, which are well described by (drifting) Kappa distribution functions rather than Maxwellian \citep{Leubner2004b, Pierrard2010}. In the fast solar wind drift speed of the proton beams is comparable or larger than Alfv{\'e}n speed $v_A$ (vary between 1.0 and 2.5 $v_A$) \citep{Marsch1982, Tu2004}, but it may be larger than 10 $v_A$ in shocks \citep{Paschmann1981}. 

In collision-poor plasmas from space, the wave fluctuations and their interactions with plasma particles are expected to play a major role conditioning plasma properties and the evolution of velocity distributions. 
Indeed, proton beams, usually guided by the local magnetic field, have been observed in association with enhanced low-frequency electromagnetic (EM) fluctuations \citep{Hoppe1983, Le1989, Sanderson1985, Wicks2016}, suggesting that these fluctuations are produced by the ion beam instabilities. In directions parallel to the background magnetic filed ($\bm{B_0}$), i.e.\  $\bm{k}\times \bm{B_0}=0$, linear theory predicts three different EM ion beam instabilities: left-hand (LH) resonant, non-resonant, and right-hand (RH) resonant ion-ion instabilities \citep[and refs.\ therein]{Gary1993}. The LH electromagnetic ion cyclotron (EMIC) instability may also be induced when the proton beam displays temperature anisotropy with $T_{b, \perp}>T_{b, \parallel}$ \citep{Shaaban2016}. Ion-ion instabilities have been extensively investigated, but their studies are in general limited to simplified models of VDFs, see \citet[and references therein]{Gary1993}, which ignore the interplay of electrons or suprathermal populations, ubiquitous in space plasmas. 

Direct in-situ measurements unveil the existence of suprathermal populations in the solar wind \citep{Stverak2008, Pierrard2010, Pierrard2016, Lazar2017Dual}, planetary bow shocks \citep{Gosling1989, Masters2016}, interplanetary shocks \citep{Wilson2019I, Wilson2019II}, CMEs \citep{Neugebauer2013}, and cometary environments \citep{Myllys2019}. The interplay of these populations and their effects on kinetic instabilities cannot be ignored. Recent studies show that kinetic (selfgenerated) instabilities are highly dependent on the shape of particle distribution, and the presence of suprathermal populations can markedly alter the dispersion and stability of EM plasma modes, e.g., EMIC, proton firehose, and proton mirror instabilities.  \citep{Shaaban2016ApSS, Vinas2017, Shaaban2017, Shaaban2018, Shaaban2018MN, Ziebell2019}. Moreover, kinetic anisotropies of electrons and their suprathermal populations may also have a significant influence on proton instabilities \citep{Kennel1968, Michno2014, Shaaban2015,Shaaban2016ApSS, Shaaban2018,Shaaban2018MN}. 

Present analysis is intended to an extended investigation of EM ion-ion instabilities, for complex but realistic physical conditions corresponding to the observations of collision-poor plasmas from space. To do so, for the velocity distributions we adopt advanced models, which combine a Maxwellian core and a drifting Kappa beam for proton populations, and bi-Kappa distribution for anisotropic electrons. Maxwellian limits ($\kappa \to \infty$) enable us to revisit and contrast with previous results obtained for idealized plasma conditions, including drifting-Maxwellian proton beams and isotropic thermal (Maxwellian) electrons. In order to isolate the effects of suprathermal populations and keep our analysis straightforward proton temperatures are considered isotropic. The new approach enables us to study for the first time ion-ion instabilities driven by the proton (counter-)beams under the effects of anisotropic electrons and suprathermal, proton or electron populations. 
In the next, our paper is organized as a detailed parametric analysis, as follows. In section \ref{sec.2}, we introduce the above-mentioned velocity distribution models for plasma particles and derive the dispersion and stability formalism on the basis of kinetic (Vlasov-Maxwell) equations. The EM ion-ion instabilities are discussed in detail in Sections ~\ref{sec.3}--\ref{sec.5}, analysing the effects of the beam velocity, suprathermal populations present in the proton beams, electron temperature anisotropies, and suprathermal electrons. The results of the present work are summarized and discussed in Section~\ref{sec.6}

\section{Dispersion relation}\label{sec.2}

We consider a three-component plasma consisting of two counter-drifting proton populations and non-drifting electrons (subscript $j=e$ in the next). Protons consist of a relatively tenuous beam (subscript $j=b$) and a dense core (subscript $j=c$)
\begin{align}
f_p=\delta f_b+ \eta f_c,
\label{e1}
\end{align}
where $\delta=~n_b/n_e$ and $\eta=1-\delta$ are the relative densities for the beam and core, respectively, and $n_0\equiv n_e$ is the total number density. The core component is assumed well described by a drifting bi-Maxwellian \citep{Shaaban2018}
\begin{align}
f_{c}\left( v_{\parallel},v_{\perp }\right) =&\frac{\pi
^{-3/2}}{\alpha_{\perp c}^{2} ~ \alpha_{\parallel c}}\exp \left[
-\frac{v_{\perp}^{2}}{\alpha_{\perp c}^{2}}-\frac{\left(v_{\parallel }-V_c\right)^{2}} {\alpha_{\parallel c}^{2}}\right],
\label{e2}  
\end{align}
with thermal velocities $\alpha_{\perp, \parallel}$ defined in terms of the corresponding temperature components, perpendicular ($T_\perp$) and parallel ($T_\parallel$) to the background magnetic filed:
\begin{align}
T_{\perp c}=\frac{m_p ~\alpha_{\perp c}^2}{2 k_B},~\text{and} ~
T_{\parallel c}=\frac{m_p ~\alpha_{\parallel c}^2}{2 k_B}.
\label{e3}
\end{align}
The beam component can be assumed a drifting bi-Maxwellian, similar to Eq. \eqref{e3}, or more general as a drifting bi-kappa distributed \citep{Shaaban2018}
\begin{eqnarray} \label{e4}
f_{b}\left( v_{\parallel},v_{\perp}\right)=&&\frac{1}{\pi ^{3/2} \theta_{\perp b}^{2}~ \theta_{\parallel b}}
\frac{\Gamma\left( \kappa +1\right)}{\kappa^{3/2}\Gamma \left( \kappa -1/2\right)}\nonumber\\
&&\times \left[ 1+\frac{v_{\perp}^{2}}{\kappa~\theta_{\perp b}^{2}}+\frac{(v_{\parallel}-V_b)^{2}}{\kappa~\theta_{ \parallel b }^{2}}\right]^{-\kappa-1},
\end{eqnarray}
with parameters $\theta_{\perp, \parallel b}=\sqrt{\left(2-3/\kappa\right)k_B T_{\perp, \parallel b}/m_p}$ defined in terms of the temperature components.
Here $V_j$ is the drifting velocity, either for the proton beam (subscript "$j=b$") or core (subscript "$j=c$"), while non-drifting electrons are assumed bi-Maxwellian or bi-Kappa distributed, as obtained from Eqs. \eqref{e3} or \eqref{e4} without drifts, i.e., $V_e=0$. We perform our analysis in a charge quasi-neutral electron-proton plasma $n_e\approx n_p= n_c+n_b$ with zero net current ensured by the proton counterbeams $n_c V_c+n_b V_b = 0$.

For a collisionless and homogeneous plasma the linear kinetic dispersion relations derived for the electromagnetic modes propagating in directions parallel to the background magnetic field (${\bm B}_0$), i.e., ${\bm k} \times {\bm B}_0 = 0$, reads \citep{Shaaban2019ApJ}
\begin{eqnarray}\label{e5}
 \dfrac{c^2 k^2}{\omega^2} = 1  +
\sum_{j=c,b,e} \dfrac{\omega_{p, j}^2}{\omega^2} \int d\bm{v} \dfrac{v_\perp} {2\left(\omega - k v_{\parallel} \mp \Omega_j\right)} \nonumber&&\\
\times\left[(\omega - k v_{\parallel}) \dfrac{\partial f_{j}}{\partial v_{\perp}} + 
k v_{\perp} \dfrac{\partial f_{j}}{\partial v_{\parallel}} \right],
\end{eqnarray}
where  $c$ is the speed of light, $k$ is the wave number, $\omega_{p, j}= \sqrt{4\pi n_j e^2/m_j}$ and $\Omega_j=e B_0/m_j c$ are, respectively, the non-relativistic plasma frequency and the gyro-frequency of species $j$, and $\mp$ differentiate between, respectively, the left-handed (LH) and  right-handed (RH) circular polarization. 
We may rewrite this dispersion relation in terms of the normalized plasma quantities, as 

\begin{align}
\tilde{k}^2&=\nonumber\\
&\delta\left[\left(A_b-1\right) +\frac{A_b\left(\tilde{\omega}\mp 1-\tilde{k}v_b \right)\pm 1}{\tilde{k}\sqrt{\beta_c}} Z_\kappa\left(\frac{\tilde{\omega}\mp 1-\tilde{k}v_b}{\tilde{k}\sqrt{\beta_b}}  \right) \right]\nonumber\\
&+\eta\left[\left(A_c-1\right) +\frac{A_c\left(\tilde{\omega}\mp 1+\tilde{k}v_c \right)\pm 1}{\tilde{k}\sqrt{\beta_c}} Z\left(\frac{\tilde{\omega}\mp 1-\tilde{k}v_c}{\tilde{k}\sqrt{\beta_c}}  \right) \right]\nonumber\\
&+\mu\left(A_e-1\right) +\mu\frac{A_e\left(\tilde{\omega}\pm \mu \right)\mp \mu}{\tilde{k}\sqrt{\beta_e}} Z_\kappa\left(\frac{\tilde{\omega}\pm \mu}{\tilde{k}\sqrt{\beta_e}}\right),
\label{e6}
\end{align}
where $\tilde{k}=ck/\omega_{p,p}$, is the normalized wave-number, $\tilde{\omega}=~\omega/\Omega_p$ is the normalized wave frequency, $\mu=~m_p/m_e$ is the proton-electron mass ratio, $\beta_j\equiv 8\pi n_0 T_{\parallel, j}/B_0^2$ is the parallel plasma beta for the component $j$, $v_j=V_j/v_A$ and $A_j=T_\perp/T_\parallel$ are their drift velocities and temperature anisotropies, respectively, 
\begin{equation}\label{e7}
Z\left( \xi _c^{\mp }\right) =\frac{1}{\pi ^{1/2}}\int_{-\infty
}^{\infty }\frac{\exp \left( -x^{2}\right) }{x-\xi _{c}^{\mp }}dx,\ \
\Im \left( \xi _{c}^{\mp }\right) >0,
\end{equation}
is the plasma dispersion function \citep{Fried1961} of argument $\xi_{c}^{\mp}=\left(\omega \mp \Omega _{p}-kV_c\right)/k\alpha_{\parallel c}$, and 
\begin{align}
Z_\kappa\left(\xi _j^{\mp }\right) =&\frac{1}{\pi ^{1/2}\kappa ^{1/2}}\dfrac{\Gamma(\kappa)}{\Gamma(\kappa-1/2)}\nonumber\\
&\times\int_{-\infty
}^{\infty }\frac{\left( 1+x^{2}/\kappa\right)^{-\kappa} }{x-\xi _{j}^{\mp }}dx, ~~\Im \left( \xi _{j}^{\mp }\right) >0,
 \label{e8}
\end{align}
is the generalized modified dispersion function \citep{Lazar2008} of argument $\xi_{b}^{\mp}=\left(\omega \mp \Omega _{p}-kV_b\right)/k\theta_{\parallel b}$ for proton beams, and $\xi _{e}^{\pm}=\left(\omega \pm \Omega _{e}\right)/k\theta_{\parallel e}$ for electrons.

The default plasma parameters used in our numerical analysis are summarized in Table~\ref{t1}, unless otherwise specified.
%
%
\begin{table}[b]
	\centering
	\caption{Default plasma parameters used in our analysis}
   \label{t1}
	\begin{tabular}{lccc} 
		\hline
		 & Beam ($b$)  & Core ($c$) &  Electrons ($e$)\\
		\hline
		$n_j/n_e$  & 0.05 & 0.95 & 1.0\\
		$T_{j,\parallel}/T_{e,\parallel}$ & 10.0 & 1.0 & 1.0\\
		$m_j/m_e$ & 1.0 & 1.0 & 1/1836 \\
		$\kappa-$index & $1.6-\infty$ & $\infty$ &  $1.8-\infty$\\
		$A_{j}$ & 1.0 & 1.0 & $\gtreqless$ 1.0 \\
		\hline
	\end{tabular}
\end{table}
%
%

\section{EMIC-beaming  instability}\label{sec.3}
%

\begin{figure}[t]
\centering
\includegraphics[scale=0.5, trim=3.3cm 3.7cm 2.9cm 3.cm, clip]{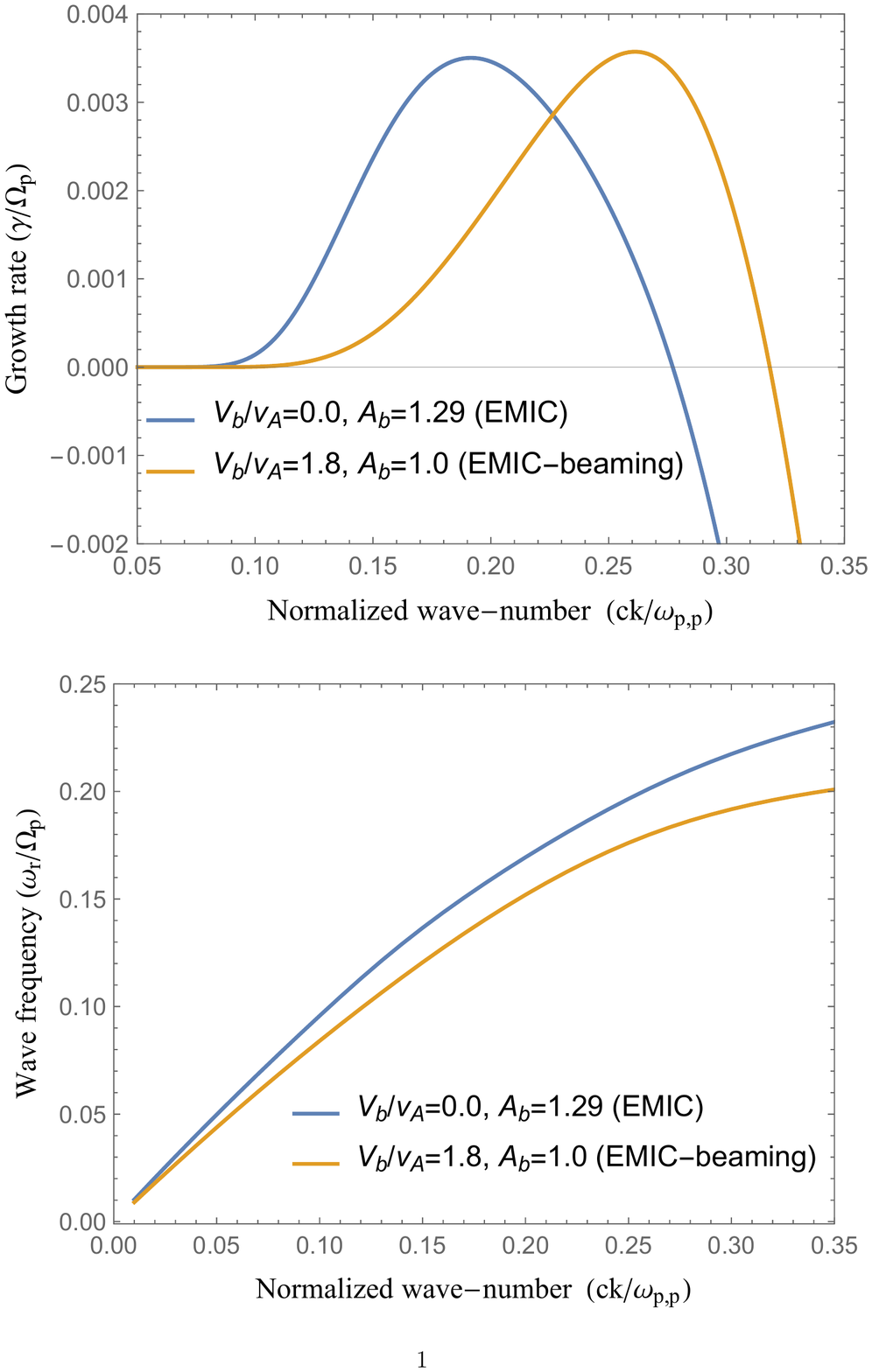}
\caption{A comparison between the growth rates (top) and wave frequencies (bottom) for EMIC and EMIC-beaming instabilities driven by $A_b=1.29$ and $V_b=1.8v_A$, respectively. Other parameters are $\delta=0.05$, $\kappa\rightarrow \infty$ $\beta_e=~\beta_c=1.0$ and $T_b=10T_c$.}
\label{fig1}
\end{figure}

\begin{figure}[t]
\centering
\includegraphics[scale=0.5, trim=3.3cm 3.7cm 2.9cm 3.cm, clip]{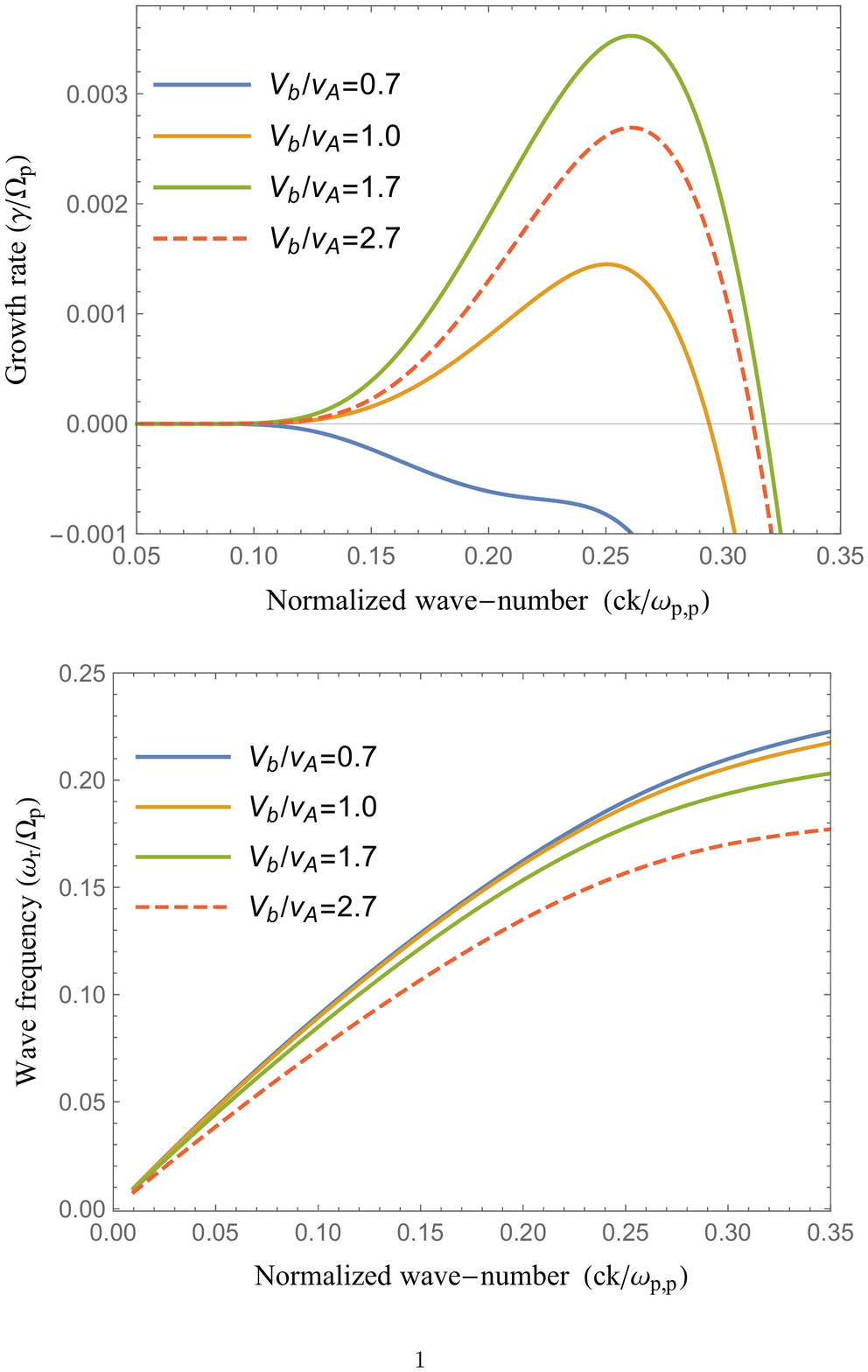}
\caption{The variation of the growth rate of the EMIC-beaming instability as a function of the beam protons drift velocity $V_b/v_A=0.7, 1.0, 1.7, 2.7$. Other plasma parameters are $\delta=0.05, \kappa=\kappa_e\rightarrow \infty, A_j=1, \beta_e=\beta_c=1.0$ and $T_b=10T_c$.}
\label{fig2}
\end{figure}
\begin{figure}[t]
\centering
\includegraphics[scale=0.52, clip]{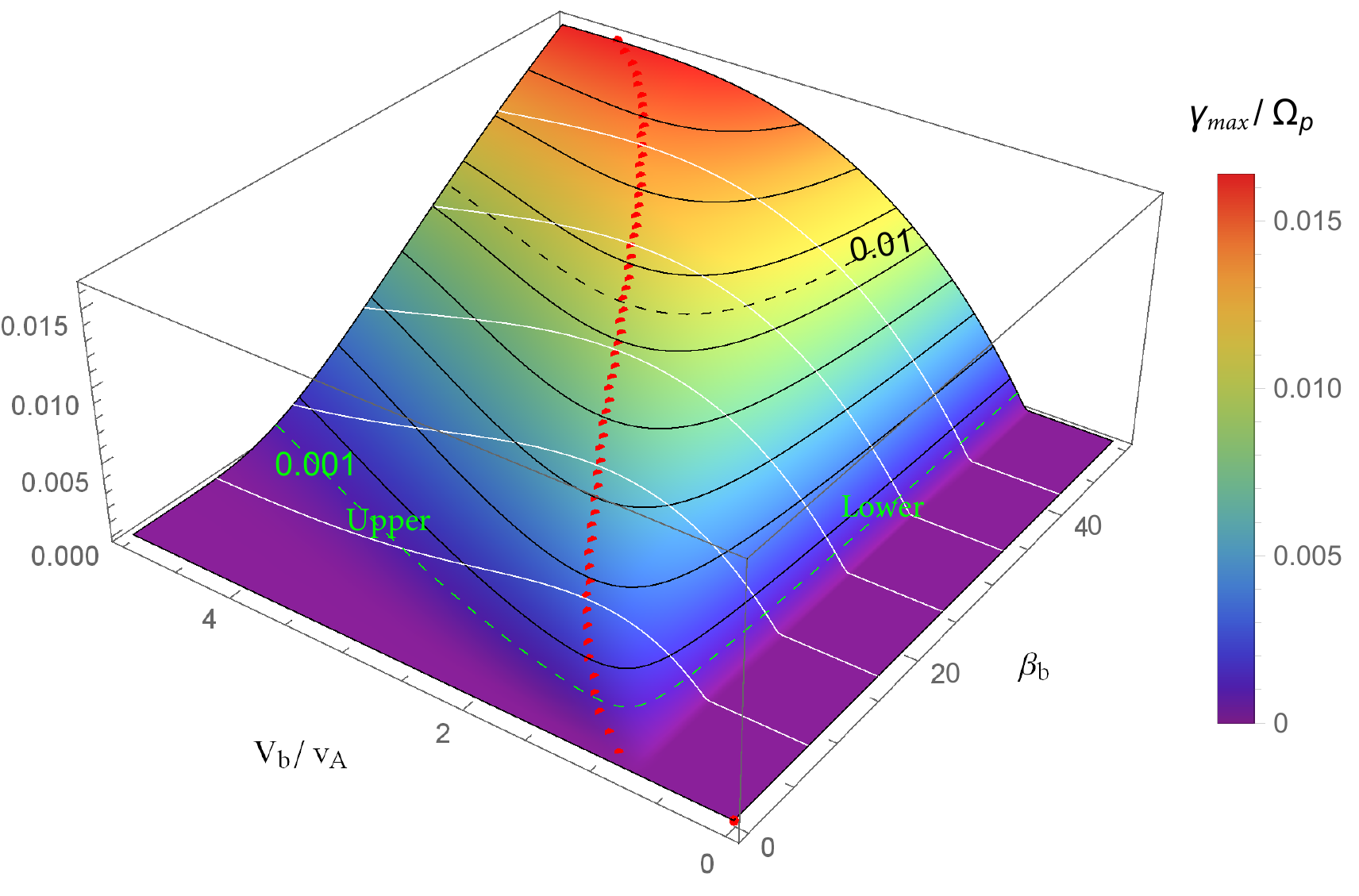}
\includegraphics[scale=0.39]{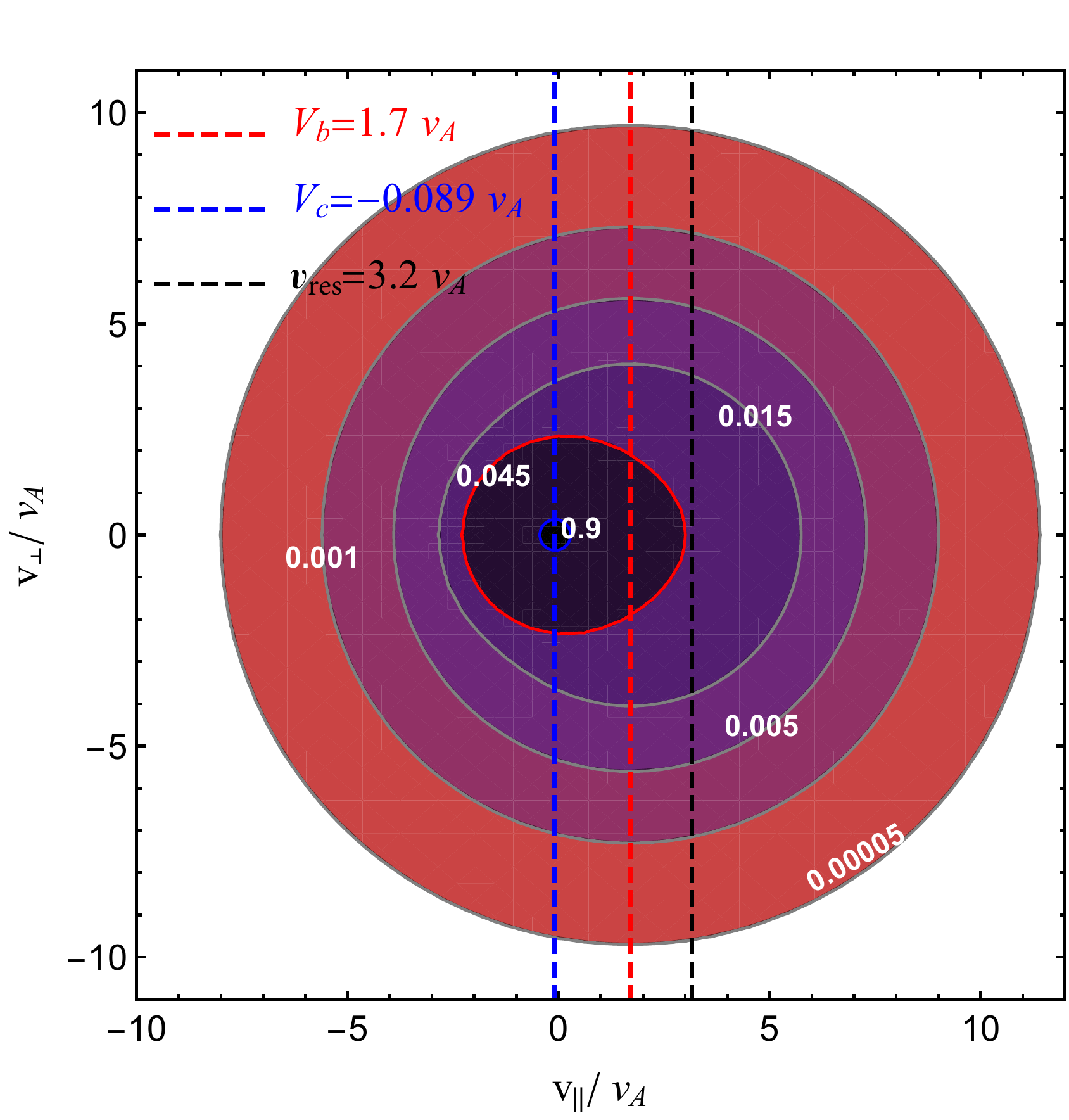}
\caption{Top panel: variation of the EMIC-beaming growth rates in terms of the beam velocity $V_b/v_A=[0-5]$ and beam plasma beta $\beta_b=[2-50]$. Bottom panel: proton counterbeams VDF susceptible to the EMIC-beaming instability in Figure~\ref{fig2} for $V_b=1.7 v_A$.}
\label{fig3}
\end{figure}
\begin{figure}[t]
\centering
\includegraphics[scale=0.52, trim=2.65cm 1.9cm 1.8cm 3.cm, clip]{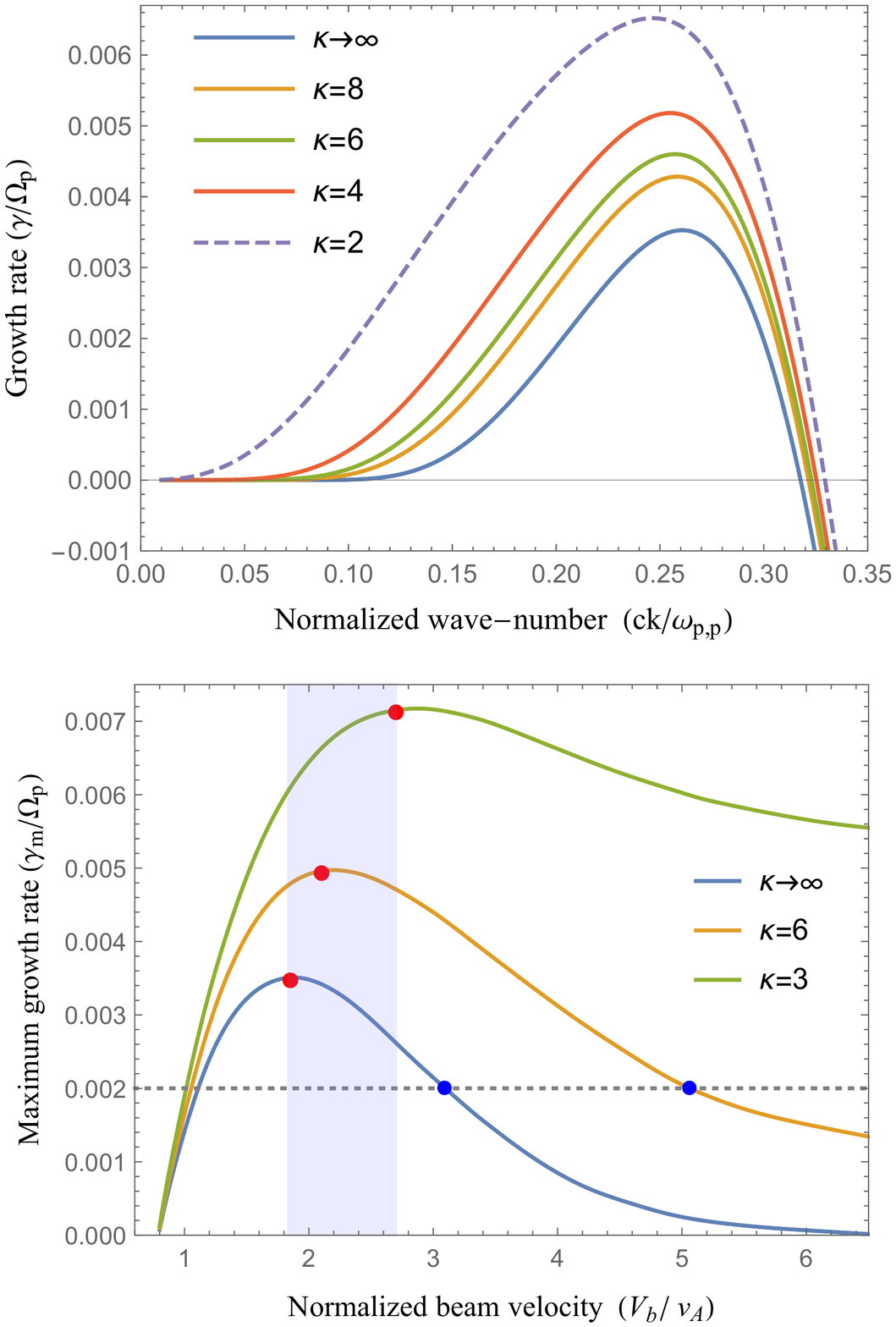}
\caption{Effects of the beam protons $\kappa-$index on the EMIC-beaming instability: growth rate in the top panel, and the maximum growth rates as a function of $V_b/v_A$ in the bottom panel. Other plasma parameters are $\delta=0.05, A_j=~1, \kappa_e\rightarrow\infty, \beta_e=\beta_c=1.0$ and $T_b=10T_c$.}
\label{fig4}
\end{figure}   

Surprisingly but true, the EM ion (proton) cyclotron (EMIC) instability can be driven by counter-drifting proton populations, for conditions typically experienced in space plasmas \citep{Gary1993}. The unstable solutions for EMIC modes can be derived numerically from the LH dispersion relation \eqref{e6}, i.e.,  for $\xi_p^-$ and $\xi_e^+$. We start our analysis with the idealized case of proton beams modeled by drifting Maxwellians ($\kappa\rightarrow \infty$ in equation \eqref{e8} for the proton beam), which is however a commonly used approach, see \cite{Gary1993} and references therein. Later we will consider the beam described more generally by a drifting Kappa (or drifting bi-Kappa) distribution function, which accounts for the presence of suprathermals. The analysis adopts the same steps for the electrons, first assuming them Maxwellian and then Kappa distributed \citep{Stverak2008, Yu2018}.

In order to motivate our interest for the unstable LH ion/ion modes, Figure~\ref{fig1} provides a comparison between the dispersive characteristics of the drift-driven LH ion/ion mode and the temperature anisotropy-driven EMIC instabilities, the growth rates (top) and the wave frequencies (bottom). The LH ion/ion instability (orange lines) is driven by counterbeaming protons with isotropic temperatures ($A_{b,c}=1$) with drift velocity $V_b=1.8 v_A$, where $v_A=B_0/\sqrt{4 \pi\ n_0 m_p}$ is the proton Alfv{\'e}n speed. Instead, the EMIC instability (blue lines) is driven by a non-drifting beam ($V_b=0$, this could the case of the so-called suprathermal halo, which is also a central component, less dense but much hotter than the core), with the same parameters, but with an intrinsic temperature anisotropy $A_b=1.29$. Figure~\ref{fig1} demonstrates that for certain conditions both EMIC and LH ion/ion instability develop along the direction of the background magnetic field ($\bm{k}\times \bm{B}_0=0$) with LH polarization, and display comparable growth rates (top panel), unstable wave-number range, and wave frequencies (bottom panel). It becomes clear to us that EMIC and LH ion/ion instabilities share the same dispersion characteristics, and therefore we call the last EMIC-beaming instability. These instabilities may develop for similar conditions, and in the presence of both sources of free energies, i.e., temperature anisotropy and drift speeds, it becomes difficult to identify the operative regimes of each instability. Therefore, in the rest of our study we assume proton beams with isotropic temperatures, which allows us to isolate the EMIC-beaming instability from any interplay with the standard EMIC instability.

Figures~\ref{fig2} presents EMIC-beaming solutions for different beaming velocities $v_b\equiv V_b/v_A=0.7, 1.0, 1.7, 2.7$, but for the same relative density $\delta=~0.05$, plasma beta parameters $\beta_e=\beta_c=1.0$, and beam-core temperature ratio $T_b/T_c=10$. Growth rates show a non-uniform variation with the beam velocity $v_b$, increasing and then decreasing with increasing $v_b$, see the dashed line for $v_b=2.7$. The corresponding wave frequencies (bottom panel) decrease with $v_b$. Similar behavior was found for the whistler heat-flux instability driven by the counter-beaming electrons \citep{Shaaban2018MN, Shaaban2018PoP}.

Physical conditions required for maximum growth rates can be identified in  Figure~\ref{fig3}, top panel, which displays maximum growth rates of EMIC-beaming instability as functions of beam velocity $V_b/v_A$ and plasma beta $\beta_b$. Shown with dashed lines are the low-level contours $\gamma_{\rm max}/\Omega_p = 0.001$ and 0.01, approaching marginal stability ($\gamma_{\rm max} \to 0$) and usually considered as thresholds of the instability; notice the resemblance with thresholds of whistler heat flux instability driven by electron beams \citep{Shaaban2018MN, Shaaban2019QLWHF}. Between upper and lower thresholds these maxima reach peaking values which increase with increasing $V_b/v_A$ and $\beta_b$, along the dotted red line. For a given $\beta_b$ the maximum growth rate increases starting from low drifts $V_b$ up to the red-line crest, and then decreases again. The explanation (similar to whistler heat-flux instability) is given by an effective temperature anisotropy that is induced by the beaming population in perpendicular direction ($T_\perp > T_\parallel$) {\it only} for intermediary values of the drift or beaming velocity. 
This anisotropy becomes apparent within high-energy (low-level) contours of VDF, as shown in Figure~\ref{fig3}, bottom panel. Here we consider the same counter-beaming protons ($V_b=1.7 v_A$, $\beta_c=1$ and $\beta_b=10$) as in Figure~\ref{fig1} and Figure~\ref{fig3} (top panel), with velocities in parallel and perpendicular directions normalized to the proton Alfv{\'e}n speed ($v_A$). 
Relative drift velocities of the beam and core components are marked with dashed red and blue lines, respectively, and black line shows resonant velocities $v_{\rm res} = |\omega_{\rm max}-\Omega_p|/k_{\rm max}$. 

\subsection{Effects of suprathermal protons}

Moving to more realistic scenarios in space plamas, we now assume the proton beam as a drifting bi-Kappa. Figure~\ref{fig4} shows the effects of suprathermal beaming protons on the EMIC-beaming instability, driven by $V_b/v_A=1.7$ for different beam protons power index $\kappa=2, 4, 6, 8$ and $\infty$ (the abundance of suprathermals increases with decreasing $\kappa$). The growth rates (top panel) markedly increase with increasing the presence of the suprathermal protons in the beam (decreasing $\kappa$). For $\kappa=2$ the maximum growth rate is two times higher than that for a Maxwellian beam ($\kappa \rightarrow \infty$). The corresponding wave frequencies remain almost constant and are not shown here. Bottom panel of Figure~\ref{fig4} displays the maximum growth rate $\gamma_m/\Omega_p$ as a function of the beam velocity $V_b/v_A$ for different $\kappa=3, 6$ and $\infty$. Similar to Figure \ref{fig3}, top panel, but for a fixed $\beta_c=1$, the maximum growth rates ($\gamma_m/\Omega_p$) of EMIC-beaming instability increases and then decreases as a function of $V_b/v_A$. Maximum growth rates are markedly enhanced with increasing the presence of suprathermal protons (lowering $\kappa$). For $\kappa=3$ the fastest growing mode is found for $V_b=2.7~v_A$ with a maximum growth rate two times greater than that obtained for $\kappa \rightarrow \infty$ for $V_b\approx1.8~v_A$. It is clear that the beam velocity required for the EMIC-beaming instability to display maximum growth rate is increasing with decreasing $\kappa$ (physical explanations are provided later in Figure~\ref{fig9}), see the purple shaded area. Consequently, the existence of the unstable EMIC-beaming modes is extended to higher drift velocities, 1.6 times higher for $\kappa=6$ than that found for the Maxwellian proton beams ($\kappa \rightarrow \infty$) before reaching the quasi-stable states below the instability threshold of maximum growth rate $\gamma_m/\Omega_p=2\times 10^{-3}$, see the dashed gray line. It is worth noticing that for high beaming speeds $V_b > 6.5 v_A$ the EMIC-beaming mode may still be unstable (with $\gamma_m/\Omega_p>0$) for $\kappa=3$ and $6$, but we cannot distinguish between the EMIC-beaming instability and another dominant ion/ion non-resonant mode, as the one discussed below in Section~\ref{sec.4}.

\begin{figure}[t]
\centering
\includegraphics[scale=0.51, trim=3.3cm 3.7cm 2.85cm 3.cm, clip]{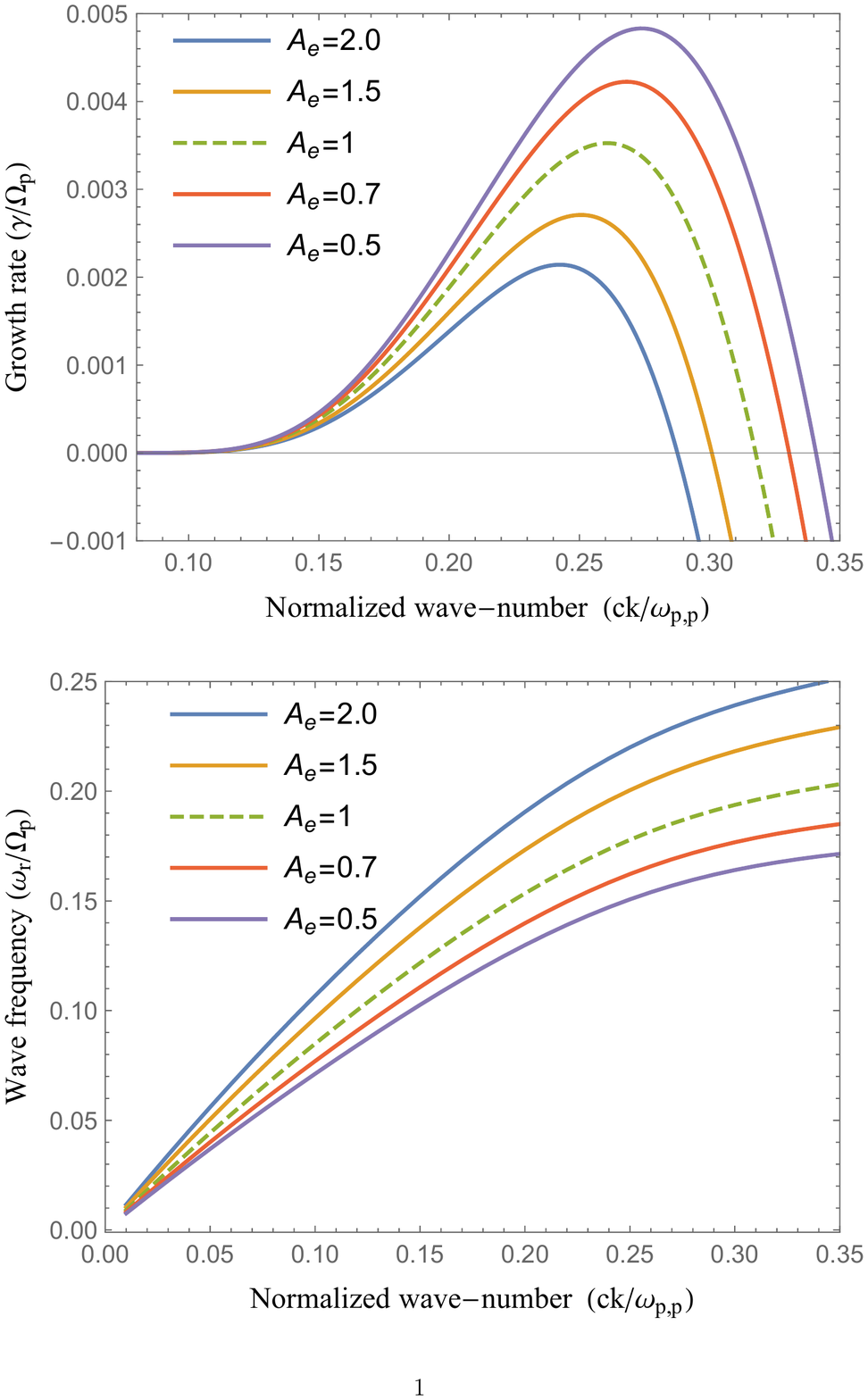}
\caption{Effects of the electron anisotropies $A_b\neq1$ on the growth rate (top) and wave frequency (bottom) of the EMIC-beaming instability.  Other plasma parameters are the same as Figure~\ref{fig2}.}
\label{fig5}
\end{figure}
%

%
\subsection{Effects of electrons}

Recently \cite{Shaaban2016ApSS, Shaaban2017} have studied the temperature anisotropy-driven EMIC instability showing that their main properties are markedly altered by anisotropic electrons and their suprathermal populations. Motivated by these results, here we discuss the EMIC-beaming instability. 
\begin{figure*}[t]
\centering
\includegraphics[scale=0.9, trim=3.cm 13.8cm 1.3cm 2.5cm, clip]{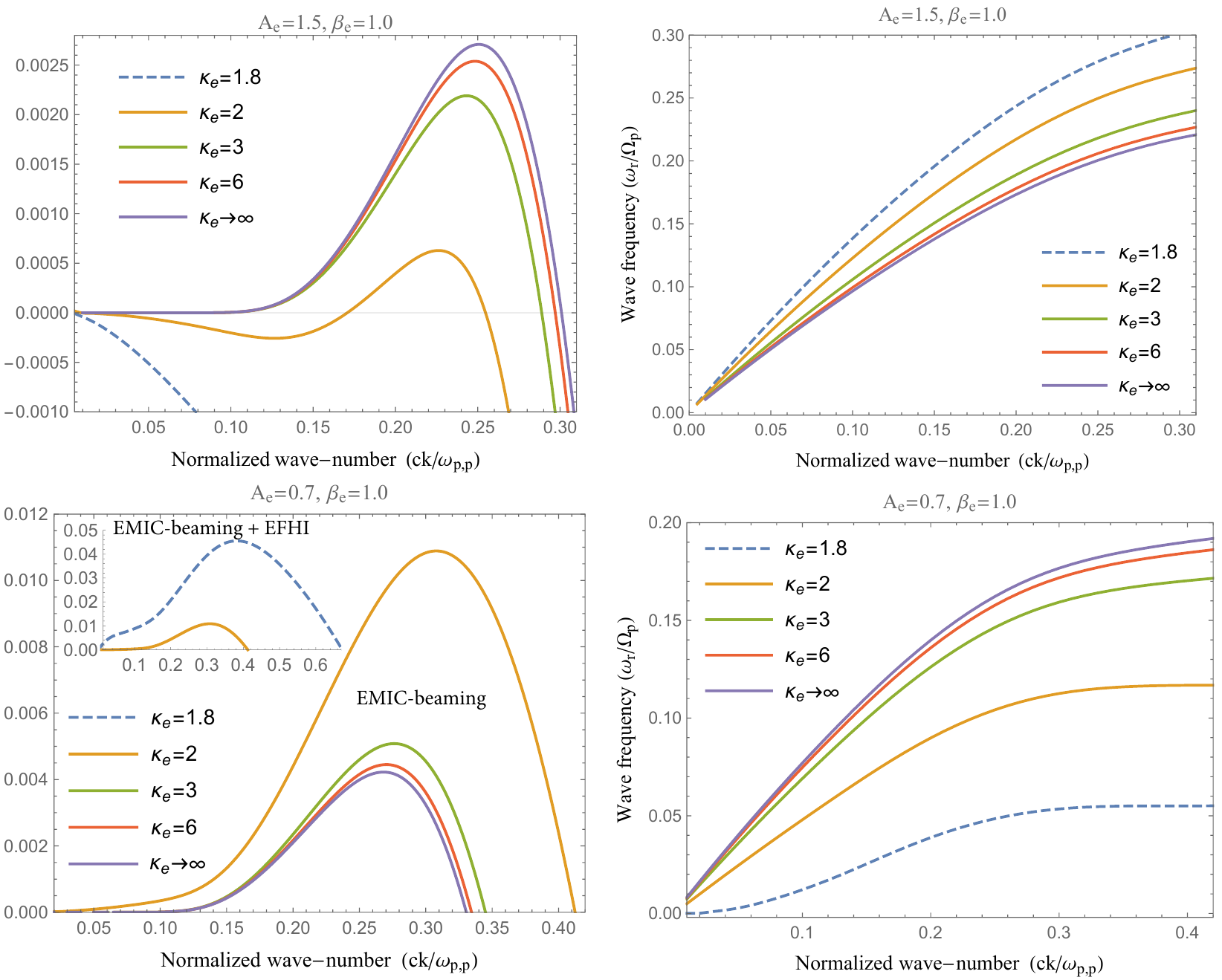}
\caption{The variation of the growth rate of the EMIC-beaming instability as a function of power-index $\kappa_e$ of the anisotropic electrons $A_e=1.5$ (top) and 0.7 (bottom). Other plasma parameters are $V_b=1.7 v_A$, $ \delta=0.05, A_{b,c}=1, \beta_e=\beta_c=1.0$, $T_b=10T_c$, and  $\kappa\rightarrow\infty$.}
\label{fig6}
\end{figure*}
Figure~\ref{fig5} describes the effects of anisotropic bi-Maxwellian electrons with $A_e=2.0, 1.5, 1.0, 0.7, 0.5$ on the  growth rates (top panel) and wave frequencies (bottom panel) of EMIC-beaming instability driven by $v_b=1.7$, and other parameters as in Figure~\ref{fig1}. Serving as reference, the growth rate and wave frequency for isotropic electrons, i.e., $A_e=1$, are displayed with dashed green line. The electron anisotropies  $A_e>1$ have inhibiting effects on the EMIC-beaming instability, decreasing the growth rates and the range of unstable wave numbers with increasing the electron anisotropy in perpendicular direction. On the other hand, the electrons with opposite anisotropy $A_e<1$ stimulate the instability, by increasing the growth rates and the range of unstable wave numbers. Contrary to growth rates the wave-frequency increases with increasing the electron anisotropy in perpendicular direction ($A_e>1$), and decreases with increasing the electron anisotropy in parallel direction ($A_e<1$).  

Figure~\ref{fig6} displays the growth rates (left panels) and wave frequencies (right panels) of the EMIC-beaming instability under the influence of anisotropic bi-Kappa electrons with $A_e=1.5$ (top panels) and $A_e=0.7$ (bottom panels). Increasing the presence of the suprathermal electrons can markedly boost their effects on this instability, see the orange lines for $\kappa_e=2$, when the maximum growth rate obtained for $A_e=1.5$ is two times lower than that for bi-Maxwellian electrons ($\kappa\rightarrow \infty$) (top panel), while the growth rate associated with $A_e=0.7$ is two times larger than that for $\kappa\rightarrow \infty$ (bottom panel). Note that anisotropies are high enough to trigger whistler instability (WI, for $A_e>1$) or electron firehose instability (EFHI, for $A_e<1$). However, we observe that high-frequency effect of WI cannot easily interfere with the low-frequency EMIC-beaming instability \citep{Shaaban2017}, while the selected plasma beta $\beta_{e, \parallel}=1$ is not sufficiently high to trigger the EFHI. Lower values of $\kappa_e \to 1.5 $ lead to significant deviations from the other unstable solutions. For instance, for $\kappa_e=1.8$ the EMIC-beaming becomes damped for $A_e=1.5$, see the dashed blue line in the top panel. In the opposite situation, when electrons exhibit an excess of parallel temperature $A_e=0.7$, small electron kappa indices, i.e., $\kappa_e=1.8$, may induce an EFHI with significantly high growth rates even for low $\beta_e\leqslant1$ \citep{Lazar2017fh}. Thus, the unstable solutions derived here for $\kappa_e=1.8$ cumulate both the EMIC-beaming and EFH instabilities, from an interplay of beaming protons $V_b=1.7 v_A$ with anisotropic suprathermal electrons $A_e=0.7$, see the dashed blue line in the bottom panel. These cumulative growth rates associated with $\kappa=1.8$ are about five times larger than that for $\kappa_e=2$, see the built-in figure in the bottom panel.

\begin{figure}[t]
\centering
\includegraphics[scale=0.5, trim=3.3cm 3.4cm 2.8cm 2.6cm, clip]{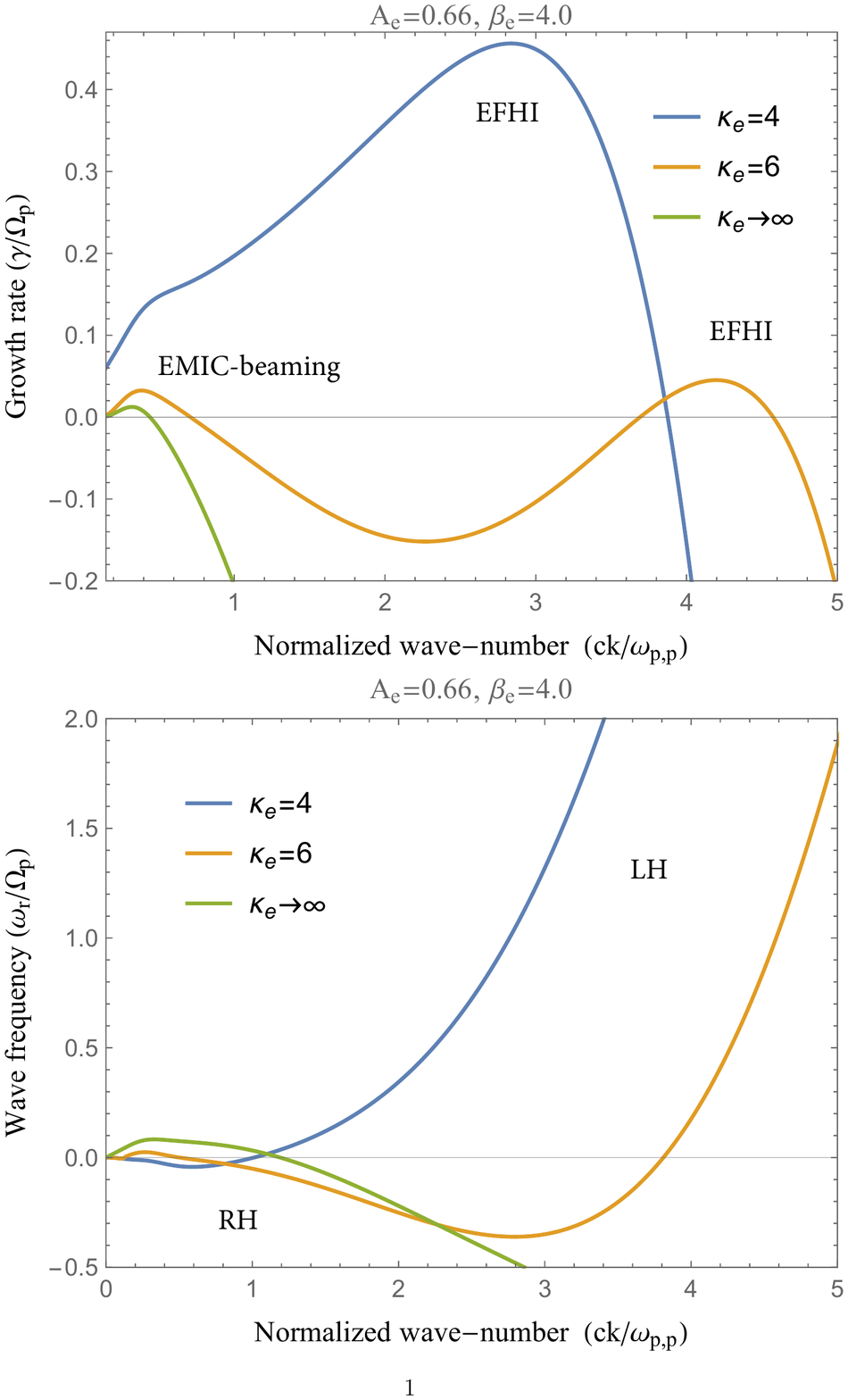}
\caption{Growth rates (top) and wave frequencies (bottom) of the EMIC-beaming and EFH instabilities driven by drift velocity of beam protons $V_b/v_A=1.7$ and electron anisotropy $A_e=0.66$, respectively, for $\beta_{\parallel e}=4$ and $\kappa_e=4, 6$, and $\infty$. Other plasma parameters are the same as Figure~\ref{fig6} }
\label{fig7}
\end{figure}

In Figure~\ref{fig7} we adopt a higher value $\beta_{e, \parallel}=4$ to study the interplay of the EMIC-beaming instability driven by $V_b/v_A=1.7$ and the EFHI driven by temperature anisotropy $A_e=0.66$ for different values of the electron power-index $\kappa=4, 6$ and $\infty$. For $\kappa\rightarrow \infty$ the growth rate displays only one peak for the EMIC-beaming instability, while for $\kappa=6$ the growth rate displays two distinct peaks corresponding to the EMIC-beaming instability (first peak at low wave numbers) and EFHI (second peak at larger wave numbers). With increasing the abundance of suprathermal electrons the EFHI becomes dominant and we cannot determine the growth rate of the EMIC-beaming instability. The stimulation of EFHI in the presence of suprathermal electrons has been quantified in recent works \citep{Lazar2017fh, Shaaban2019a, Lopez2019a}. Bottom panel shows LH wave frequencies ($\omega_r>0$) in the range of the unstable wave numbers, corresponding to the EMIC-beaming and EFHI instabilities. Out of this range the wave frequency is RH polarized ($\omega_r<0$) and corresponds to damped whistler modes.

%
\section{Ion-Ion non-resonant instability}\label{sec.4}
%
\begin{figure}[t]
\centering
\includegraphics[scale=0.7, trim=5.2cm 9.9cm 4.6cm 3.cm, clip]{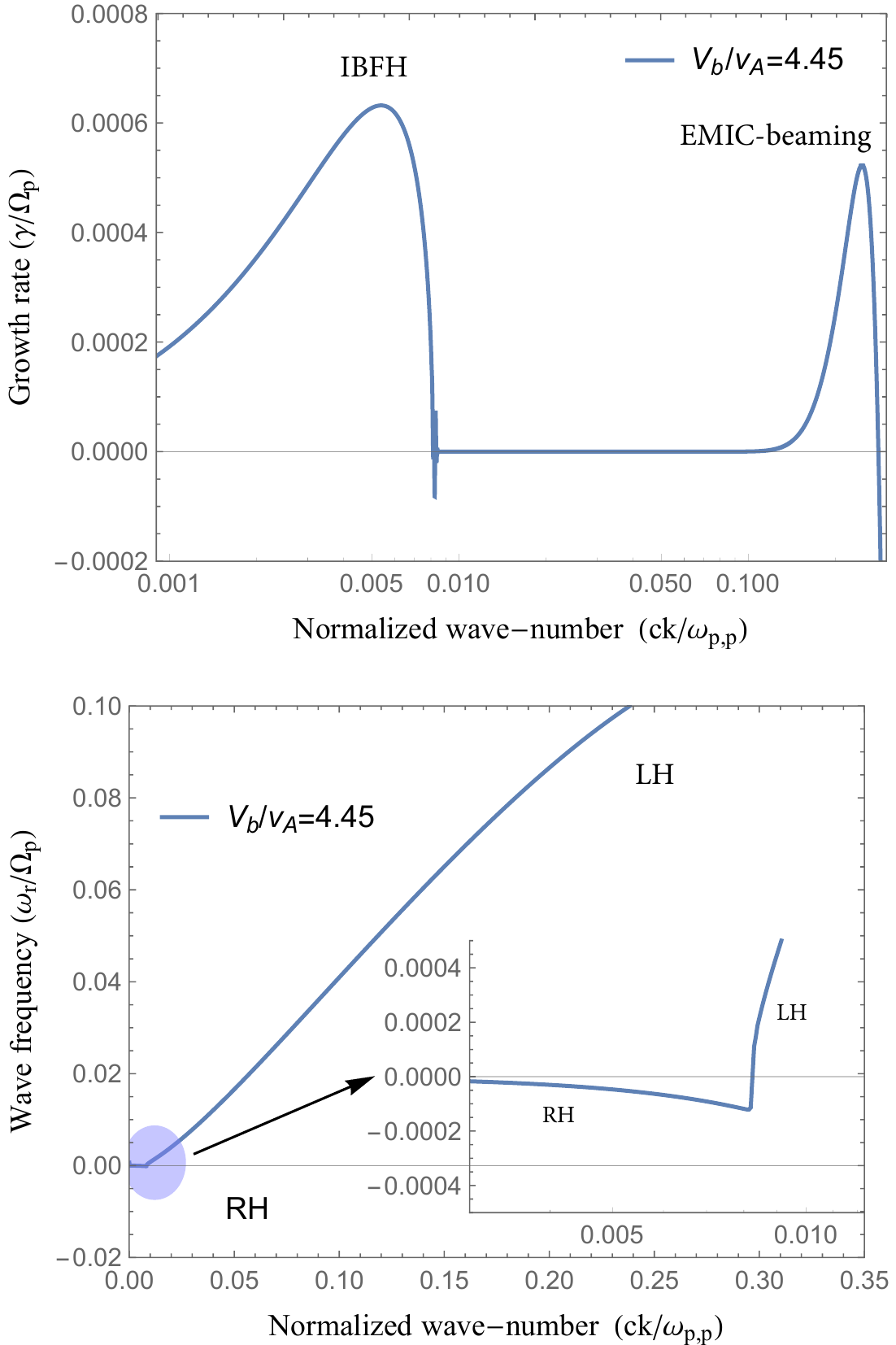}
\includegraphics[scale=0.45]{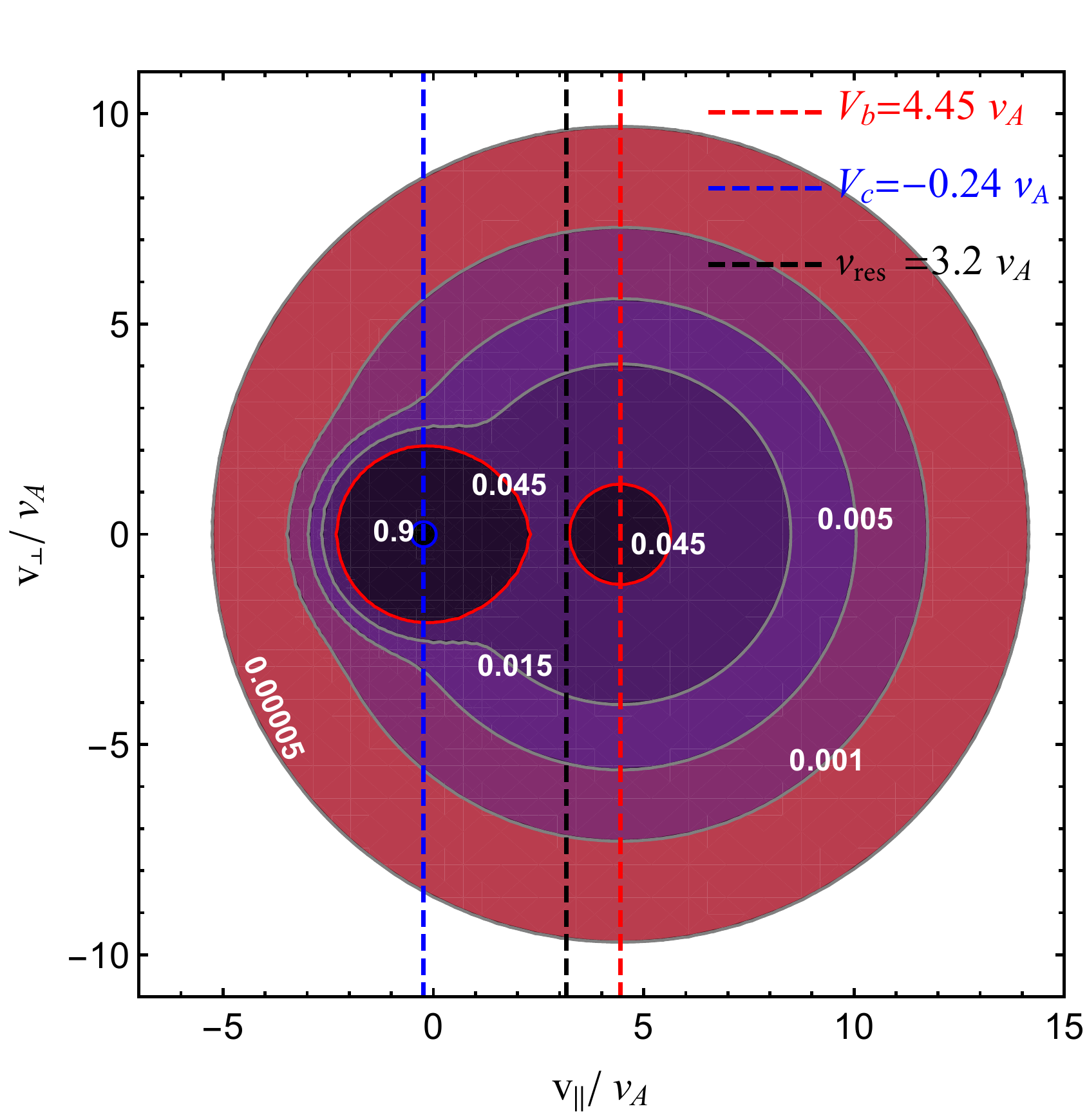}
\caption{Transition between the LH EMIC-beaming and the RH IBFH instabilities at beaming velocity $V_b=4.45v_A$: growth rate (top), wave frequency (middle), and the corresponding proton counterbeams VDF (bottom). Other plasma parameters are $\delta=0.05, A_j=~1, \kappa=\kappa_e\rightarrow\infty, \beta_e=\beta_c=1.0$ and $T_b=10T_c$. }
\label{fig8}
\end{figure}

\begin{figure*}[t]
\centering
\includegraphics[scale=1, trim=3.cm 14.4cm 2.5cm 2.8cm, clip]{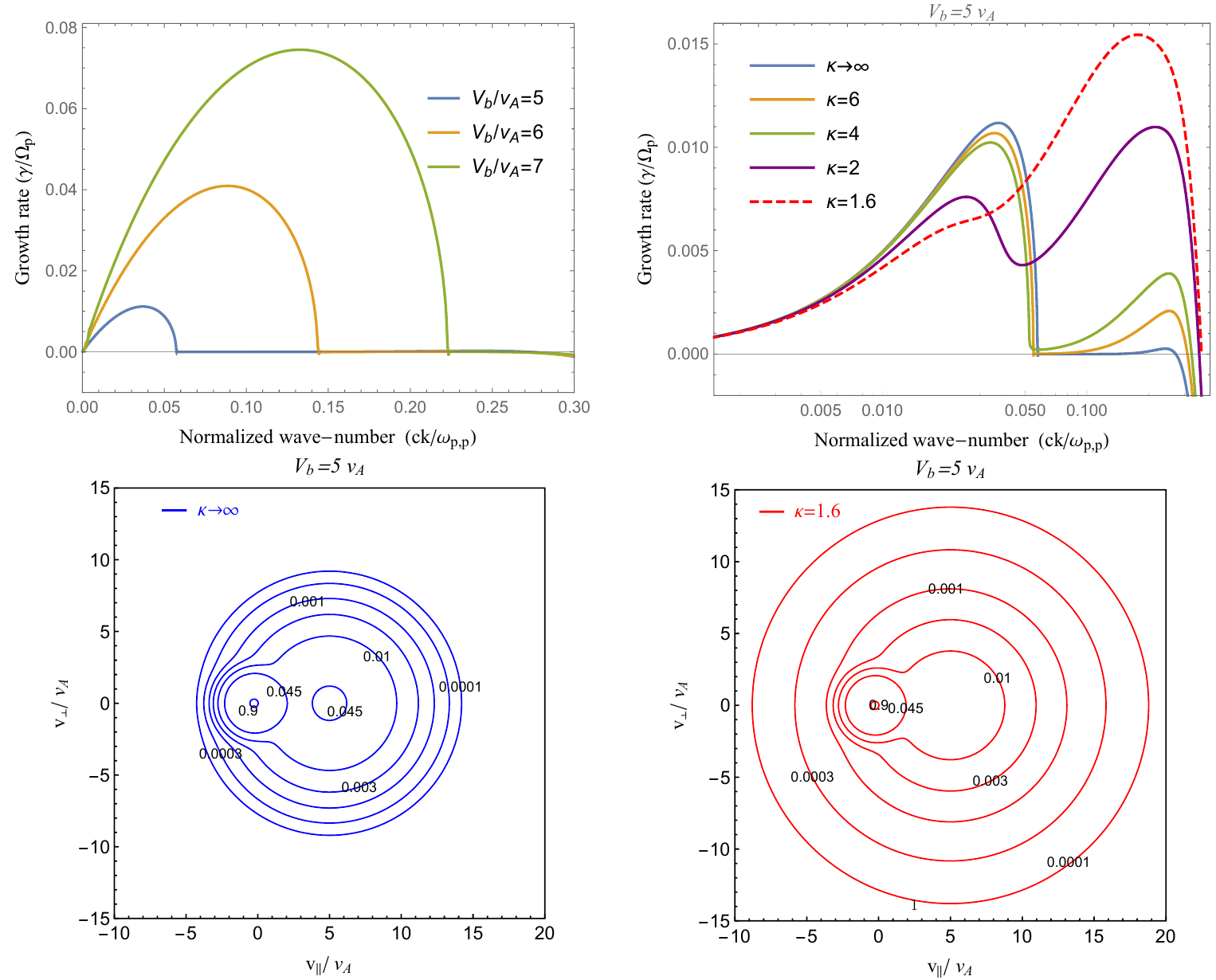}\caption{Top left: effects of the beaming velocity $V_b/v_A=5,6,7$ of Maxwellian beam ($\kappa\rightarrow\infty$) on the growth rate of IBFH instability. Top right: effects of the suprathermal protons ($\kappa=1.6,2,4,6, \infty$) on IBFH and EMIC-beaming instabilities driven by $V_b/v_A=5$. Bottom panels display the corresponding proton counterbeams VDF. Other plasma parameters are the same as Figure~\ref{fig8}.}
\label{fig9}
\end{figure*}
The non-uniform variation of the EMIC-beaming instability as a function of beam velocity $V_b$ (with maximum growth-rate peaking for a certain value of $V_b$) directs us to another important question on what is constraining the more energetic beams, with higher beaming velocities. 
In this section we propose to answer to this question. Figure~\ref{fig8} displays the growth rate (top panel) and corresponding wave frequency (middle panel) of the ion/ion instabilities for the same plasma conditions as in Figure~\ref{fig1}, but for a higher beam velocity, i.e.,  $V_b/v_A=4.45$. The growth rate displays two distinct peaks, the first peak at low wave numbers corresponding to Ion/Ion non-resonant instability, which we name ion beaming firehose (IBFH) instability, while the second peak at higher wave numbers corresponds to the EMIC-beaming instability. In the middle panel the corresponding wave frequency confirms the transition from the  IBFH to the EMIC-beaming instabilities in terms of the mode polarization, that starts with RH polarization ($\omega_r<0$ for IBFH), and then coverts to LH polarization ($\omega_r>0$) corresponding to EMIC-beaming instability. Similar transition between the unstable LH and RH modes has been reported by \cite{Shaaban2018MN} for the electron beaming instabilities, see figure~4 therein. Bottom panel presents the proton counter-beaming distributions corresponding to the unstable modes driven by $V_b/v_A=4.45$. Here, it is obvious that the beam and core populations are weakly coupled, see the red contour of level 0.045, and the beaming velocity in parallel direction induces an effective parallel anisotropy (like $T_{\parallel, eff}>T_\perp$), which must be favorable to an RH IBFH instability. Unlike the distribution in Figure~\ref{fig3}, here the drift velocity of the beam is larger than thermal velocity of the resonant protons $V_b\approx 1.39~v_{res}$, see the dashed black line.    

\begin{figure*}[t]
\centering
\includegraphics[scale=0.68, trim=7cm 0.45cm 6.5cm 2.5cm, angle =-90 , clip]{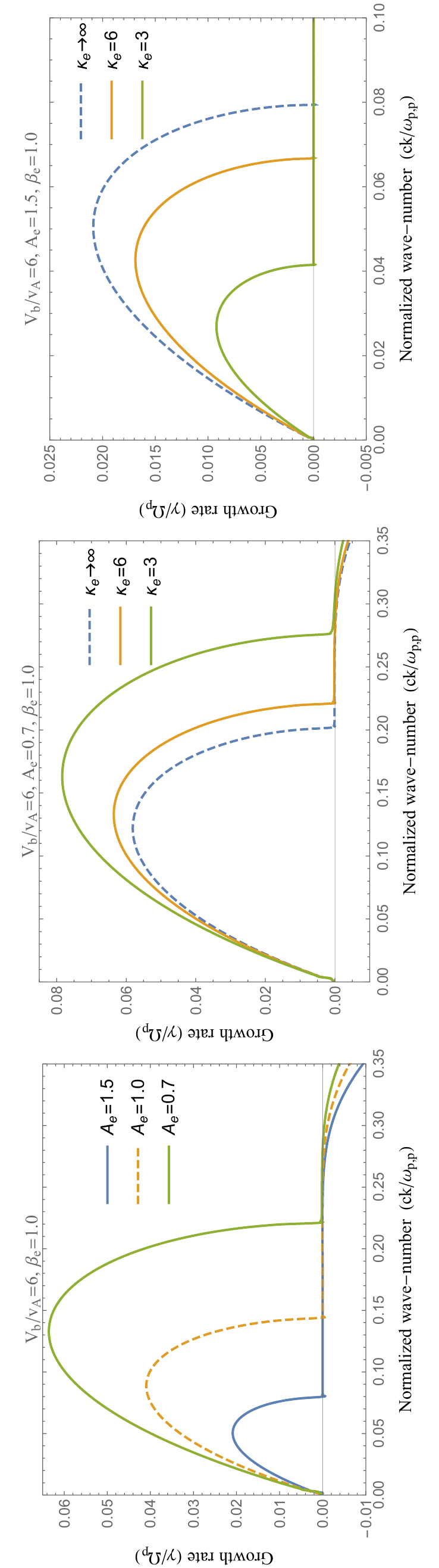}
\caption{Effects of the anisotropic electrons $A_e=0.7, 1.0, 1.5$ (left) and their suprathermal population $\kappa_e=3,6,\infty$ (middle and right) on the IBFH instability driven by $V_b/v_A=6$. Other plasma parameters are the same as Figure~\ref{fig8}. }
\label{fig10}
\end{figure*}
\subsection{Effects of suprathermal protons}
For more energetic proton beams, the IBFH instability is more operative and becomes dominant at lower wave-numbers. Thus, in Figure~\ref{fig9} we study the dispersive characteristics of this instability driven by proton beams for sufficiently large beaming velocities (left panels), and for different $\kappa$-values of proton beams (right panels). In the left-top panel proton beams are assumed Maxwellian distributed ($\kappa\rightarrow\infty$), with different velocities $V_b/v_A=5, 6,7$. The growth rates and the range of unstable wave numbers are systemically increased with increasing $V_b$. In the right-top panel we adopt $V_b/v_A=5$ to study the growth-rate variation with the power-index, e.g., for $\kappa=1.6, 2, 4, 6, \infty$. The abundance of suprathermal protons in the beam has inhibiting effects on the IBFH instability, lowering the growth rates with decreasing $\kappa$. Here, it is worth mentioning that EMIC-beaming mode is quasi-stable for $\kappa \rightarrow \infty$, see the blue line at large wave numbers. However, the presence of the suprathermal population in the beam destabilize the EMIC-beaming mode with considerable growth rates at large wave numbers, e.g., for $\kappa=6, 4$, while for $\kappa=1.6$ and $2$ EMIC-beaming instability becomes dominant, see the purple and dashed red lines. The corresponding wave-frequencies are not shown here, since their variations with $V_b/v_A$ and $\kappa$ are modest. Bottom panels of Figure~\ref{fig9} show the proton distributions used to derive the unstable IBFH and EMIC-beaming solutions in top panels, for a drifting-Maxwellian beam ($\kappa \rightarrow \infty$, left) and a drifting-Kappa with $\kappa=1.6$ (right) and $V_b/v_A=5$ in both cases. A direct comparison of these distributions shows that the contrast between the core and beam populations is reduced in the presence of suprathermals in a Kappa-distributed beam. As a  consequence of that the core and beam components become strongly coupled, as we can see comparing for instance the contours of level 0.045 in bottom panels. For $\kappa=1.6$ the counter-beaming distribution becomes more favorable to the unstable EMIC-beaming instability. This may explain the suppression of IBFH instability and the stimulation of EMIC-beaming instability, as well as the increase of the beaming velocity threshold $V_{b,th}$, as suggested in bottom panel of Figure~\ref{fig4}.

\subsection{Effects of electrons}

\begin{figure}[h!]
\centering
\includegraphics[scale=0.5, trim=3.3cm 3.7cm 2.8cm 2.5cm, clip]{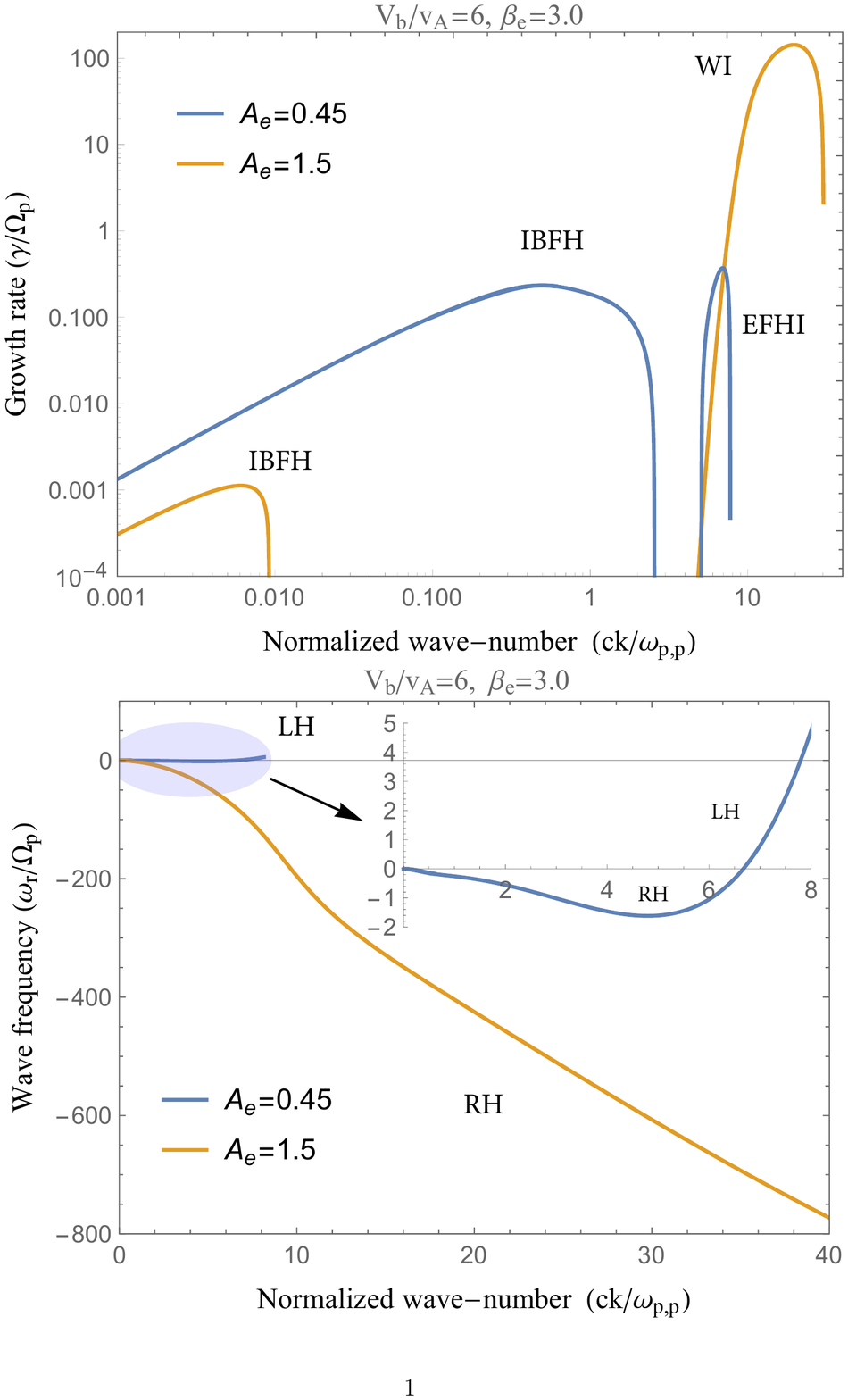}
\caption{The interplay of IBFH instability driven by $V_b=6~v_A$ with WI (driven by $A_e=1.5$)r and EFH (driven by $A_e=0.7$) for $\beta_e=3$. Other plasma parameters are the same as Figure~\ref{fig8}.}
\label{fig11}
\end{figure}

In this section we study the effects of anisotropic electrons ($A_e \ne 1$) and their suprathermals on the IBFH instability. Figure~\ref{fig10} displays growth rates obtained for $V_b=6~v_A$ and $\beta_c=1$: in left panel for $\kappa=\kappa_e\rightarrow \infty$ and different temperature anisotropies of electrons $A_e=0.7, 1.0, 1.5$, while in middle and right panels we plot the growth rates for $A_e=0.7$ and $A_e=1.5$, respectively, and different $\kappa_e=2, 6$, and $\infty$ for bi-Kappa-distributed electrons. The growth rate and the range of unstable wave numbers increases with increasing the electron anisotropy in  perpendicular direction ($A_e=1.5$, left panel), and  decreases for opposite anisotropies in parallel direction ($A_e=0.7$). All the effects of the anisotropic electrons on the IBFH instability are boosted by increasing the presence of suprathermal electrons, i.e., lowering $\kappa_e$, see middle and right panels. 

For sufficiently large plasma beta $\beta_e>1$ and electron anisotropies $A_e\neq 1$ WI and EFHI can be self-generated. Thus, in Figure~\ref{fig11} we study the interplay of the IBFH instability triggered by $V_b/v_A=6$ with the WI or EFHI driven, respectively, by $A_e=1.5$ or $A_e=0.45$, and for high beta plasma conditions $\beta_e=3~\beta_c$. The growth rates in the top panel of Figure~\ref{fig11} display four distinct peaks, two peaks for the IBFH instability at low wave numbers, a third peak peak for the EFHI at large wave numbers, and the fourth peak for the WI at larger wave numbers. In bottom panel, the corresponding wave frequencies confirm the conversion of the RH polarized IBFH to the LH EFH modes by changing the sign in between their peaks, see the zoom-in subplot. Otherwise, the RH branch of the IBFH modes at low frequencies extends smoothly to the electron scales corresponding to WI.
\begin{figure}[t]
\centering
\includegraphics[scale=0.51, trim=3.3cm 3.3cm 2.7cm 3.cm, clip]{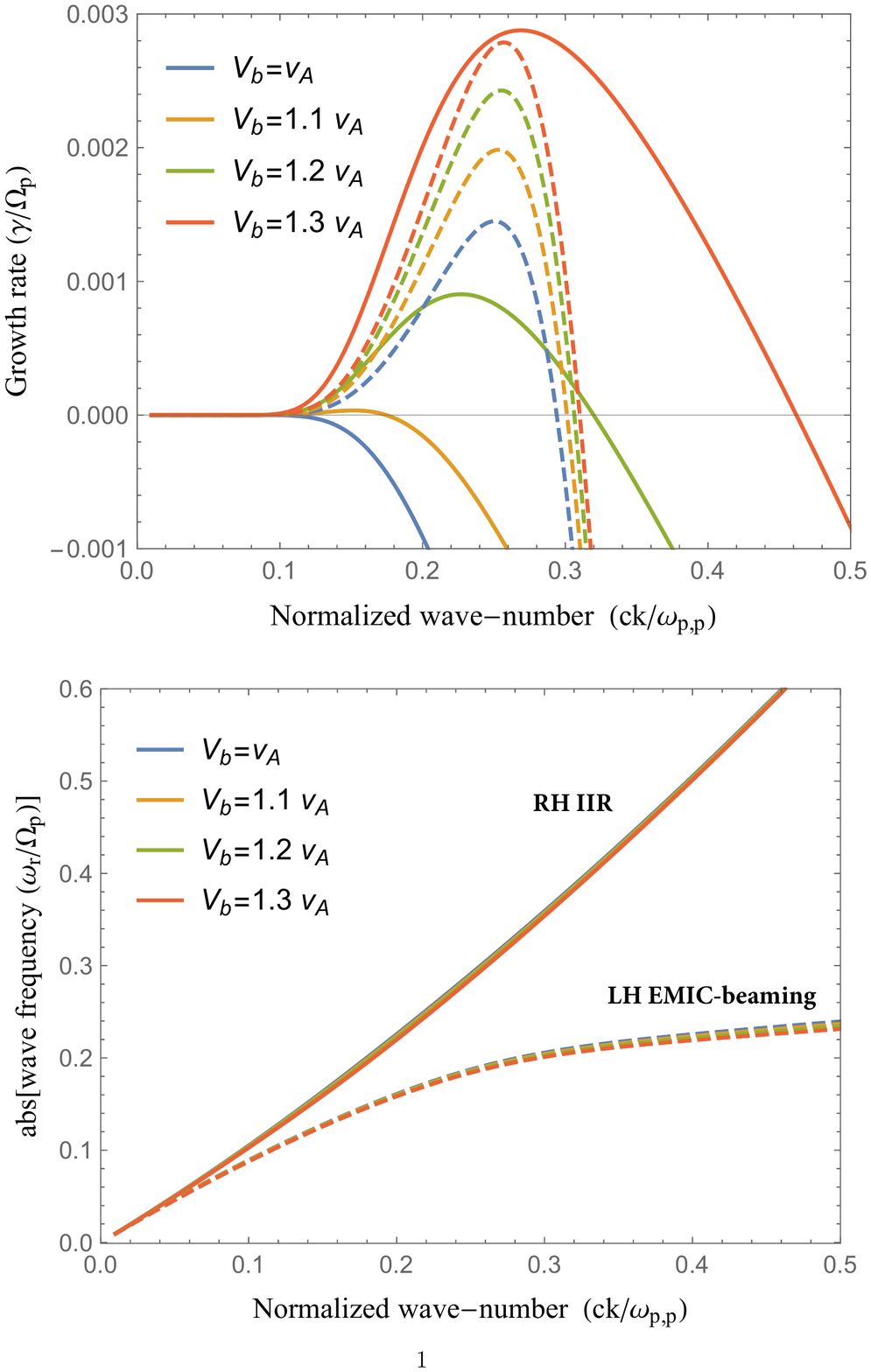}
\caption{A comparison between RH IIR instability (solid lines) and LH EMIC-beaming instability (dashed lines): Growth rates (top) and absolute wave frequency (bottom) as functions of beam velocity $V_b/v_A$. Other parameters are $\delta=0.05$, $\kappa=\kappa_e\rightarrow \infty$, $ A_j=1, \beta_e=~\beta_c=1.0$ and $T_b=10T_c$.}
\label{fig12}
\end{figure}
\begin{figure}[t]
\centering
\includegraphics[scale=0.52, trim=3.3cm 3.7cm 2.8cm 3.cm, clip]{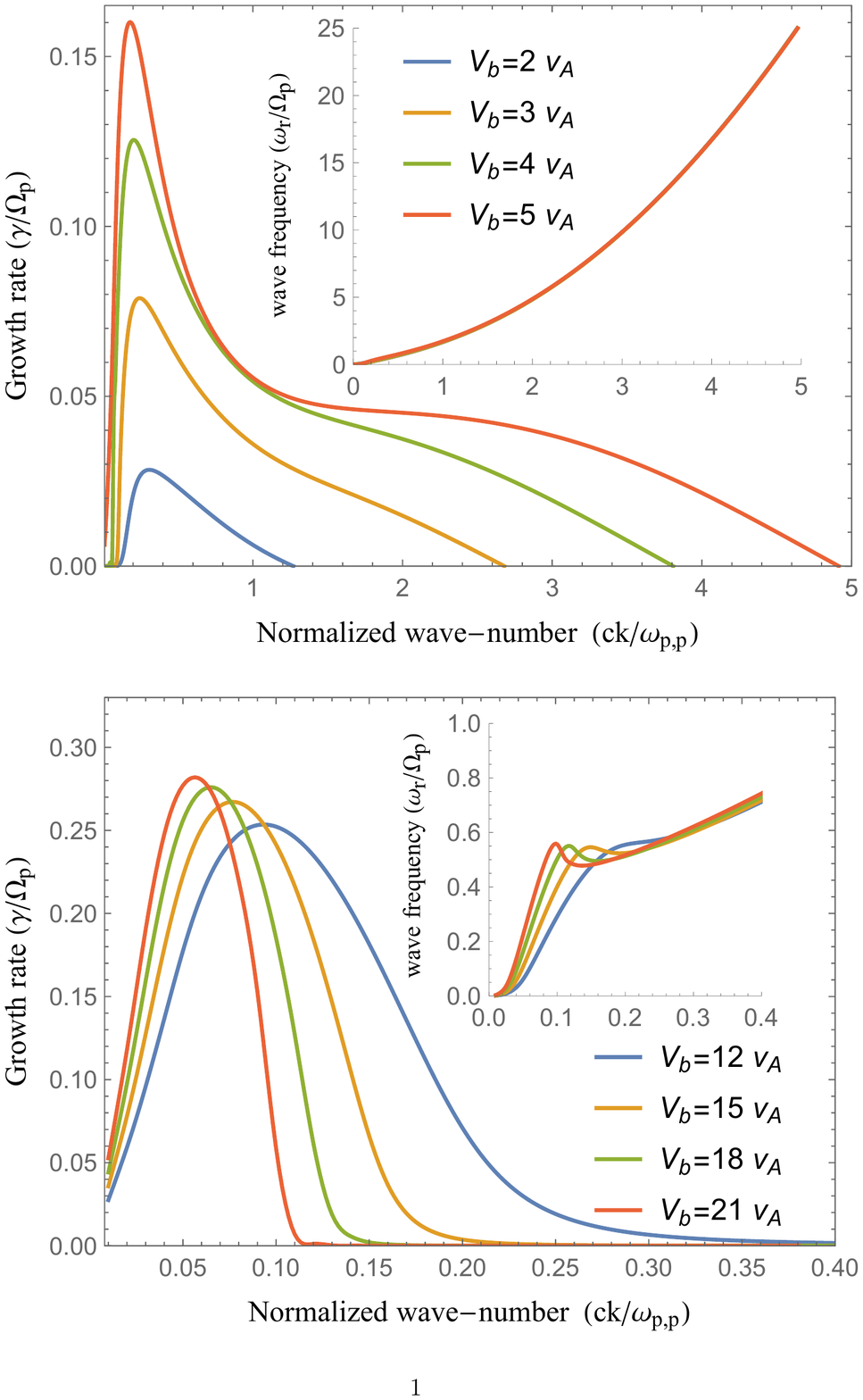}
\caption{Growth rates and wave frequencies (sub-figures) of the IIR instability for moderate drift velocities $V_b/v_A=2,3,4,5$ (top) and large drift velocities $V_b/v_A=12,15,18,21$ (bottom). Other plasma parameters are the same as Figure~\ref{fig12}.}
\label{fig13}
\end{figure}

\begin{figure*}[t]
\centering
\includegraphics[scale=0.72, trim=7.4cm 2.cm 6.7cm 2.5cm, angle =-90 , clip]{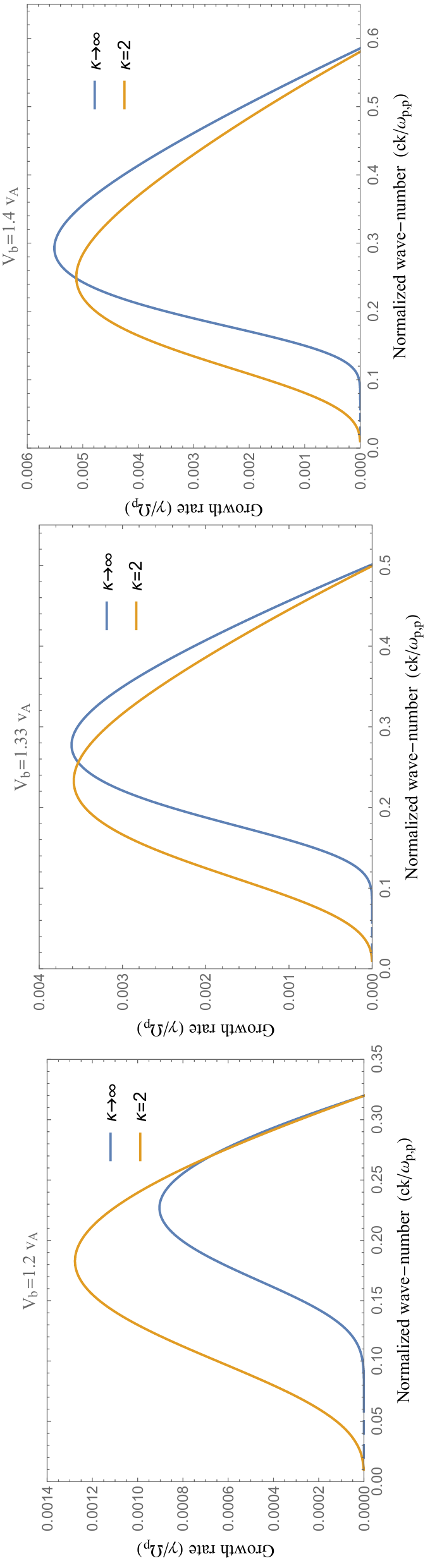}
\caption{The variation of IIR instability as a function of $\kappa$ for the beaming velocities $V_b/v_A=1.2$ (left), 1.33 (middle), and 1.4 (right). Other plasma parameters are the same as Figure~\ref{fig12}.}
\label{fig14}
\end{figure*}
%
%
%
\section{Ion-Ion RH resonant instability}\label{sec.5}
%

%
\begin{figure*}[t]
\centering
\includegraphics[scale=0.72, trim=4.07cm 2.1cm 3.5cm 2.5cm, angle =-90 , clip]{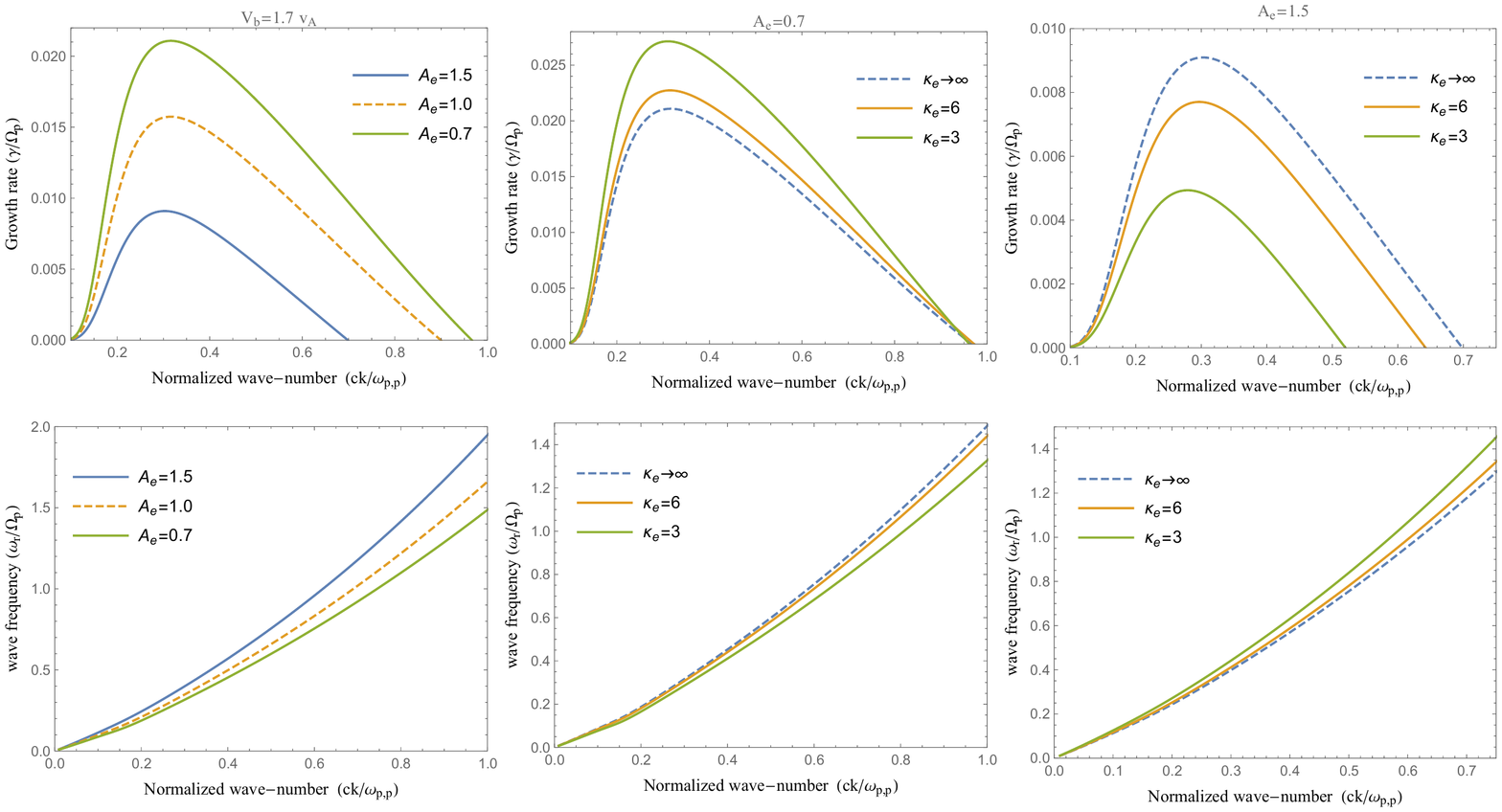}
\caption{Effects of electron anisotropy $A_e=0.7,1.0,1.5$ (left) and their suprathermal populations $\kappa_e=3, 6, \infty$ (middle and right) on the growth rates (top) and wave frequencies (bottom) of IIR instability. Other parameters are the same as Figure~\ref{fig12}.}
\label{fig15}
\end{figure*}

In a direct competition with EMIC-beaming instability is the ion-ion resonant (IIR) instability, which may develop for the same plasma conditions but with RH polarization. Here, the unstable solutions for the ion-ion modes can be derived numerically from the RH dispersion relation \eqref{e6}, i.e.,  for $\xi_p^+$ and $\xi_e^-$. Figure~\ref{fig12} shows the growth rates (top) and wave-frequencies (bottom) of the ion-ion RH resonant instability (solid lines) against those of the EMIC-beaming instability (dashed lines) for the same plasma conditions distinguished with colors. For a small beaming velocity $V_b/v_A=1.0$, the IIR mode is stable, while the EMIC-beaming mode is unstable and displays maximum growth rate of $\gamma/\Omega_p\approx1.4\times10^{-3}$. More energetic beams with $V_b/v_A=1.1$ and $1.2$ destabilize the IIR mode, but the EMIC-beaming instability remains dominant. For a higher beaming velocity $V_b/v_A\geqslant1.3$, the IIR instability becomes dominant with a growth rate exceeding that of the EMIC-beaming instability. For the sake of comparison we plot the absolute wave frequencies ($|\omega_r/\Omega_p|$) of both instabilities in the bottom panel. The wave frequencies of the unstable IIR modes are larger than those for the EMIC-beaming instability, 1.7 times larger at $\gamma_{max}-\tilde{k}$ and 3.3 times larger at $\gamma_{min}-\tilde{k}_{max}$.  Figure~\ref{fig13} presents the growths rates and the corresponding wave frequencies (sub-figures) of the IIR instability with the same conditions as in Figure~\ref{fig12}, but for larger beaming velocities $V_b \geqslant 2 v_A$. Growth rates show a uniform variation, increasing with increasing $V_b/v_A$. However, we observe two distinct operative regimes for IIR instability: first regime is conditioned by low and moderate values $1< V_b/v_A \lesssim10$, when the unstable wavenumbers increase as $V_b$ increases and the wave frequencies are in the frequency range of $\Omega_p<\omega_r\ll|\Omega_e|$, see examples in top panel for  $V_b/v_A = 2,3,4,5$. On the other hand, the second regime is conditioned by very large values $V_b/v_A>10$, in which the unstable wavenumbers decrease as $V_b$ increases and the wave frequencies are in the frequency range of $\omega_r\lesssim\Omega_p$, see examples in bottom panel for $V_b/v_A=12,15,18,21$.          
\subsection{Effects of suprathermal protons}
Particularly interesting are the unstable solutions close to the instability thresholds of low maximum growth rates, e.g., $\gamma_m=10^{-3}\Omega_p$ and $\gamma_m=10^{-2}\Omega_p$, which are obtained for low beaming (or drifting) velocities $V_b/v_A = [1.2-1.7]$. Figure~\ref{fig14} presents the effects of the suprathermal proton beams on the growth rate of the IIR instability. The growth rates are plotted for $\kappa=2$ (blue) and $\infty$ (orange) for different beam velocities $V_b/v_A=$ 1.2 (left), 1.33 (middle), and 1.4 (right). The effects of suprathermal protons highly depend on the beam velocity, enhancing the growth rate in the left panel and inhibiting them for a higher $V_b$ in the right panel, while the switch between these opposite effects occurs in this case for $V_b/v_A=1.33$ in the middle panel, where growth rates for $\kappa=2$ and $\kappa \to \infty$ are comparable. 
\subsection{Effects of electrons}
In this section we study the effects of anisotropic electrons and their suprathermal populations on the IIR instability. Left panels in Figure~\ref{fig15} show the effects of the bi-Maxwellian electrons with different temperature anisotropies, $A_e=0.7,1.0$ and $1.5$, on the growth rates (top) and wave-frequencies (bottom). Serving as a reference, the growth rate and wave frequency for isotropic electrons $A_e=1.0$ is plotted by dashed line. The growth rates are stimulated by the electron anisotropy in the parallel direction $A_e=0.7$, while inhibited by the electron anisotropy in the perpendicular direction $A_e=1.5$. The corresponding wave frequencies (bottom) show opposite behaviour, being enhanced by $A_e=1.5$, but inhibited by $A_e=0.7$. Middle and right panels show that in the presence of the suprathermal electrons the effects of anisotropic electrons on growth rates and wave-frequencies of IIR instability are markedly boosted .

\begin{figure}[t]
\centering
\includegraphics[scale=0.5, trim=3.3cm 3.7cm 2.8cm 2.6cm, clip]{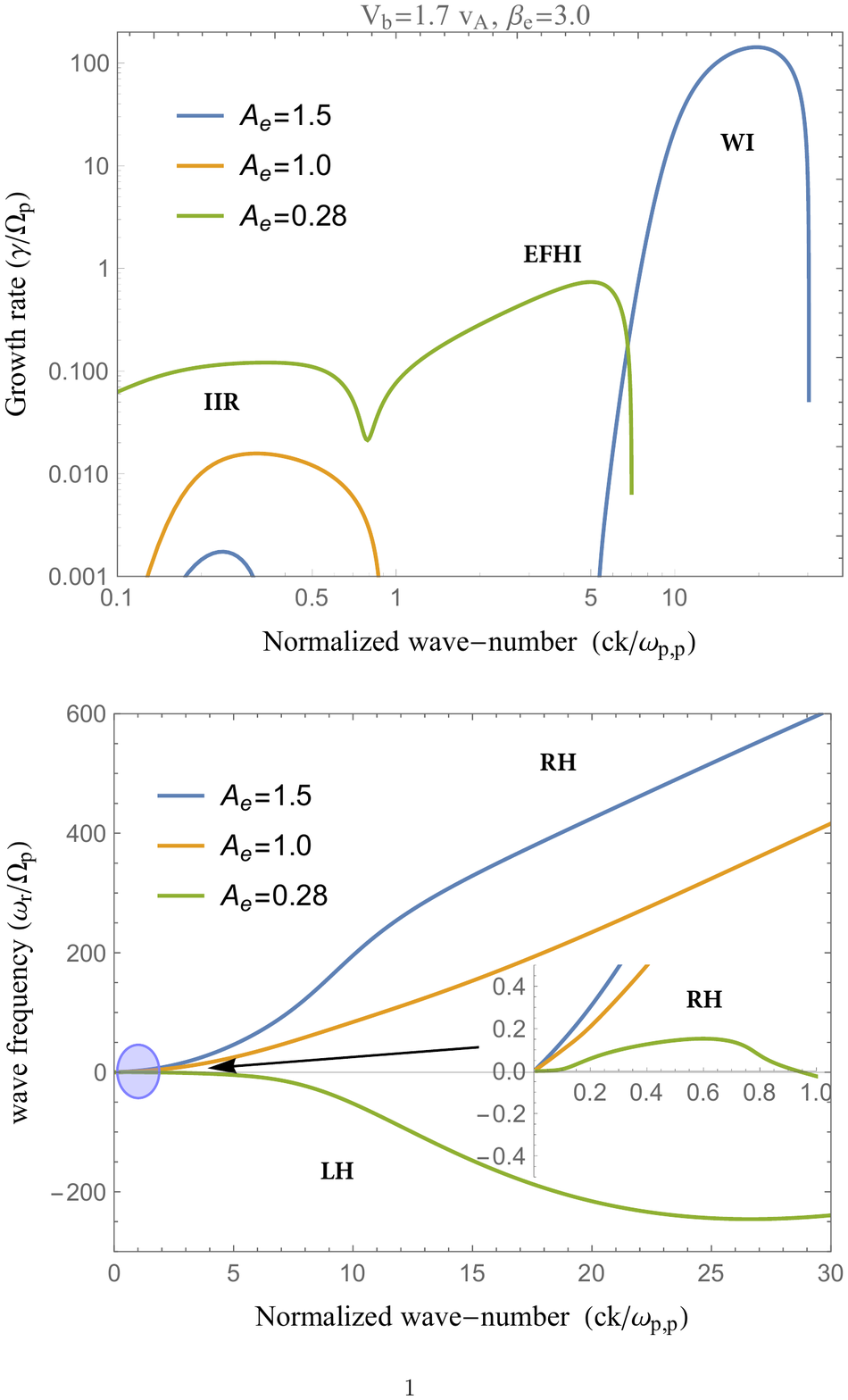}
\caption{The interplay of the IIR instability driven by beaming velocity $V_b=1.7 v_A$ with WI and EFH instabilities driven by electron anisotropies $A_e=1.5$ and 0.28, respectively, for $\beta_e=3$. Other parameters are the same as Figure~\ref{fig12}.}
\label{fig16}
\end{figure}

Finally, in Figure~\ref{fig16} we study the interplay of this instability with WI and EFHI driven by, respectively, $A_e=1.5$ and $A_e=~0.28$. In both cases $\beta_e=3.0$. Growth rates of IIR instability are located at low wave-numbers and are markedly stimulated anisotropic electrons with $A_e=~0.28$, but strongly inhibited for $A_e=1.5$. For  $A_e=0.28$ the growth rate displays a second peak at large wave-numbers corresponding to the LH EFH instability, while for $A_e=~1.5$ the growth rate displays a second peak at much larger wave-numbers corresponding to the RH WI. The corresponding wave-frequencies in bottom panel confirm our identification upon these instabilities, in different ranges of wave-frequency and with different polarizations, e.g,, $\omega_r<0$ for the LH EFHI and $\omega_r>0$ for the RH WI and RH IIR instabilities.

%
\begin{table*}[t]
	\centering
	\caption{Summary of the EM instabilities in the present paper}
   \label{t2}
	\begin{tabular}{lccccccc} 
		\hline
		Instability & Frequency & Polarization & Free energy & $\kappa_p \searrow$ & $A_e \ne 1$; ($\kappa_e \searrow$)\footnote{Increasing the presence of the suprathermal electrons (decreasing $\kappa_e \searrow$) boosts the effects of $A_e \ne 1$ on all instabilities.}\\
		\hline
		EM Ion cyclotron & $\omega_r<\Omega_p$ & Left-hand & $T_{p \perp}>T_{p \parallel}$ &  Stimulate & $A_e<1$ Stimulate\footnote{Effects of $A_e\neq 1$ on the EMIC instability can be found in \cite{Shaaban2017}.\\}\\
		 (EMIC)  &  & &  &  & $A_e>1$ Suppress\\
		     \hline
		Ion-Ion LH resonant & $\omega_r\ll\Omega_p$ & Left-hand & $V_b \lesssim v_{res}$ & Stimulate & $A_e<1$ Stimulate\\
		(EMIC-beaming)  &  & &  &  & $A_e>1$ Suppress\\
		\hline
		Ion-Ion non-resonant & $\omega_r/\Omega_p\lll 1$ & Right-hand & $V_b > v_{res}$ & Suppress & $A_e<1$ Stimulate\\
		(IBFH)  &  &  &  &  & $A_e>1$ Suppress\\	
		\hline
		Ion-Ion RH resonant & $\Omega_p<\omega_r\ll|\Omega_e|$ & Right-hand & $1<V_b/v_A\lesssim 10$ & Non-uniform & $A_e<1$ Stimulate\\
		(IIR or Magnetosonic) & $\omega_r\lesssim \Omega_p$ & & $V_b/v_A >10$   &  & $A_e>1$ Suppress\\
		\hline
	    Whistler (WI) & $\Omega_p\ll\omega_r<|\Omega_e|$ & Right-hand & $T_{e \perp}>T_{e \parallel}$ & --- & Co-exist with IIR \\
		&  &  &  &  & and IBFH\\
		\hline
		Electron firehose & $\Omega_p<\omega_r\ll|\Omega_e|$ & Left-hand & $T_{e \perp}<T_{e \parallel}$ & --- & Co-exist with IBFH, \\
		(EFHI)  &  &  &  &  & IIR and EMIC-beaming\\
		\hline
	\end{tabular}
\end{table*}
%
%
\section{Discussions and Conclusions}\label{sec.6}
In this paper we have presented a new and refined parametric analysis of the ion-ion instabilities resulted from the interplay of proton beams, anisotropic electrons ($A_e\neq 1$) and suprathermal populations of protons and electrons, focusing on conditions experienced in various space plasmas environments such as solar wind, interplanetary shocks, planetary bow shocks, coronal mass ejections, and cometary environments. Suprathermal populations are ubiquitous in space plasmas, and play a key role in our kinetic approaches, in which the proton beams and anisotropic electrons have been assumed well described by drifting Kappa and bi-Kappa distribution functions, respectively. These generalized model distributions are not only realistic, but enable direct comparisons with idealized approaches which limit to Maxwellian representations ($\kappa = \kappa_e \rightarrow \infty$), i.e., drifting-Maxwellian for proton beams, or Maxwellian for electrons. We have numerically solved the kinetic dispersion relations for the parallel EM waves, providing exact solutions for EMIC, EMIC-beaming, IBFH, and IIR instabilities, as well as WI, and EFHI instabilities.

Section~\ref{sec.3} describes the EMIC-beaming instability destabilized by the less energetic beams, a LH resonant mode with frequency $\omega_r\ll \Omega_p$ (Figure~\ref{fig1}), and growth rates conditioned by $V_b \lesssim v_{res}$ (where $V_{res}$ is the thermal velocity of resonant protons) and depending non-monotonously of the beam drift velocity (Figures~\ref{fig2} and \ref{fig3}). The maximum growth rates are derived in Figure~\ref{fig3} in terms of the beam plasma beta $\beta_b$ and beaming velocity $V_b/v_A$. Growth rates are increased by increasing $\beta_b$. Suprathermal proton beams (quantified by $\kappa$) stimulate the EMIC-beaming instability by enhancing the growth rates and extending the unstable regime to higher beaming velocities (Figure~\ref{sec.4}). EMIC-beaming instability is found to be very sensitive to the electron anisotropies $A_e\neq 1$, and their suprathermal population (quantified by $\kappa_e$): growth rates are markedly decreased if $A_e>1$, and are markedly enhanced if $A_e<1$ (Figure~\ref{fig5}). These effects are stimulated by increasing the electron suprathermal populations, see Figure~\ref{fig6} for $\kappa=2$. For a sufficiently large electron plasma beta ($\beta_e=4$), both EMIC-beaming and EFHI instabilities may co-exist and interplay if $A_e<1$ and $\kappa_e\leqslant6$ (Figure~\ref{fig7}). 

In Section~\ref{sec.4} we have studied the IBFH instability driven by more energetic beams with drift velocities $V_b/v_A> V_{res}$. In fact, $V_b/v_A\approx V_{res}$ marks a transition between the LH EMIC-beaming and the RH IBFH instability, which displays an additional peak at lower wave numbers (Figure~\ref{fig8}) and has wave frequency $\omega_r\lll \Omega_p$. This transition is physically explained by the contours of the proton distribution in Figure~\ref{fig8}-(bottom panel), which show that beaming protons become less resonant for higher beaming velocities, exciting the IBFH modes and suppressing the EMIC-beaming instability. Further increase of the beaming velocity leads to a uniform increase of growth rates and unstable wave numbers. The suprathermal protons in the beam have inhibiting effects on the IBFH instability decreasing the growth rates as $\kappa$ decreases. Physical explanations for these effects are provided in Figure~\ref{fig9}-(bottom panel) by the contours of the proton distributions, which become less favorable to this instability in the presence of suprathermal protons. The growth rates and the corresponding unstable wave-numbers of the IBFH instability are markedly increased by the electron temperature anisotropy in parallel direction, $A_e<1$, but are diminished by the electron anisotropy in perpendicular direction $A_e>1$. All these effects are markedly stimulated in the presence of the suprathermal electrons, i.e., lowering $\kappa_e$, see Figure~\ref{fig10}. For a higher electron plasma beta, e.g., $\beta_e=3$, both EFHI (driven by $A_e<1$) and WI (driven by $A_e>1$) display additional peaks at larger unstable wave numbers, suggesting that LH EFHI and RH WI instabilities can co-exist and interplay with the RH IBFH instability. Operative regimes of different instabilities have been identified by the polarization and the wave frequency ranges of the unstable modes (Figure~\ref{fig11}).

In competition with EMIC-beaming instability, the IIR instability can develope for the same plasma conditions but with RH polarization and wave frequency in the range $\Omega_p<\omega_r\ll|\Omega_e|$. A comparative analysis between the two instabilities is performed, and we have found that EMIC-beaming is dominant only for very low beaming velocities $V_b/v_A<1.3 v_A$ (Figure~\ref{fig12}). Further increase of $V_b$ enhancing the growth rates of the IIR instability, which show a uniform variation as a function of $V_b/v_A$. We have identified two distinct regimes depending on $V_b$: for low and moderate velocities $1<V_b/v_A\lesssim10$ the unstable wave numbers increase with increasing $V_b/v_A$ and the instability has wave frequency in the range $\Omega_p<\omega_r\ll|\Omega_e|$, while for large drift velocities $V_b/v_A>10$ the unstable wave numbers are decreased by increasing $V_b/v_A$ and the instability has wave frequency in the range $\omega_r\lesssim\Omega_p$ (Figure~\ref{fig13}).  The presence of the suprathermal protons leads to a non-uniform variation of the growth rates, increasing and then decreasing as $V_b/v_A$ increases, see Figure~\ref{fig14}. We have studied the effects of anisotropic electrons and their suprathermal population on IIR instability. The growth rates and the unstable wave numbers are found to be enhanced by electron anisotropy $A_e<1$, but suppressed by $A_e>1$. These effects are again stimulated in the presence of suprathermal electrons (Figure~\ref{fig15}). WI and EFHI instabilities are predicted at larger wave numbers for sufficiently large anisotropies and electron plasma $\beta_e=3$, and these instabilities may, in general, develop faster than IIR instability, see Figure~\ref{fig16}.  

We conclude stating that the present study unveils new unstable regimes for the so-called ion-ion, or ion beaming instabilities, highly conditioned by the anisotropic electrons and suprathermal populations. The instability conditions including our new results are summarized in Table~\ref{t2}. The outcomes of the present study should offer multiple and valuable explanations for the enhanced low-frequency electromagnetic fluctuations, frequently observed in association with proton beams in space plasmas.
%

\acknowledgments
The authors acknowledge support from the Katholieke Universiteit Leuven and Ruhr-University Bochum. These results were obtained in the framework of the projects 
SCHL 201/35-1 (DFG-German Research Foundation), GOA/2015-014 (KU Leuven), G0A2316N (FWO-Vlaanderen), 
and C 90347 (ESA Prodex 9). S.M. Shaaban would like to acknowledge the support by a Postdoctoral 
Fellowship (Grant No. 12Z6218N) of the Research Foundation Flanders (FWO-Belgium). R.A.L thanks the support of AFOSR grant FA9550-19-1-0384.

\bibliography{papers}

%
\listofchanges
\end{document}